\documentclass[11pt]{article}

\usepackage{amsfonts}
\usepackage{eurosym}
\usepackage{amssymb}

\usepackage{amsmath}
\usepackage{epsfig}

\usepackage[utf8]{inputenc}
\usepackage{lscape} 
\usepackage{rotating}
\usepackage{hyperref}
\hypersetup{
    colorlinks=true,
    linkcolor=blue,
    filecolor=blue,      
    urlcolor=blue,
    pdfpagemode=FullScreen,
    }

\usepackage{natbib}
\usepackage{graphicx}

\newcounter{theorem}

\newcommand{\SN}{{\rm I\kern-.23em N}}

\begin{document}
\title{The Interactions of Social Norms about Climate Change: Science, Institutions and Economics\thanks{%
We gratefully acknowledge the financial help of the European Union's Horizon 2020 research and innovation
programme under the Marie Sklodowska-Curie grant agreement No 891124.}\\
}

\author{Antonio Cabrales\thanks{Universidad Carlos III de Madrid. E-mail: antonio.cabrales@uc3m.es}, Manu García\thanks{Washington University in St. Louis. E-mail: manu@wustl.edu}, David Ramos Muñoz\thanks{Universidad Carlos III de Madrid. E-mal: dramos@der-pr.uc3m.es} \& Angel Sánchez\thanks{Universidad Carlos III de Madrid. E-mail: anxo@math.uc3m.es}} 
\date{\today. }

\maketitle

\begin{abstract}
We study the evolution of interest about climate change between different actors of the
population, and how the interest of those actors affect one another. We first document the
evolution individually, and then provide a model of cross influences between them, that we
then estimate with a VAR. We find large swings over time of said interest for the general
public by creating a Climate Change Index for Europe and the US (CCI) using news media
mentions, and little interest among economists (measured by publications in top journals
of the discipline). The general interest science journals and policymakers have a  more
steady interest, although policymakers get interested much later.

{\small \noindent \textbf{JEL Classification}: Q54, Q58, D85, A13.\newline
\noindent \textbf{Keywords}: Climate Change, Social Norms, Text Analysis, Social Networks.}
\end{abstract}

\newpage

\section{Introduction}

\emph{''The furnaces of the world are now burning about 2,000,000,000 tons of coal a year. When this is burned, uniting with oxygen, it adds about 7,000,000,000 tons of carbon dioxide to the atmosphere yearly. This tends to make the air a more effective blanket for the earth and to raise its temperature. The effect may be considerable in a few centuries.''}. August 14, 1912, Rodney \& Otamatea Times “Science Notes and News”.

As the quote above shows, the knowledge about anthropogenic climate change is not exactly new. Nor is the knowledge that it can create problems for humanity. \cite{huntington1917climatic} in the \emph{Quarterly Journal of Economics} already claimed that climate change (not necessarily anthropogenic in this case) partially explained the fall of Rome.

Tackling this problem requires that regulators of different sorts take
decisions that provide incentives for abatement. But as the references above show they are being very slow in doing this. The science about climate change has been there for a long time. So why does it seem that action is not happening sufficiently quickly? 

A first answer is that there is already some action. Many
regulators are aware of the problem. The European Commission has a Technical
Expert Group on sustainable finance (TEG) which has produced several
reports, for example, an EU taxonomy -- to determine whether an economic
activity is environmentally sustainable; an EU Green Bond Standard;
methodologies for EU climate benchmarks and disclosures for benchmarks; and
guidance to improve corporate disclosure of climate-related information. All
of this suggests that perhaps in the future we will have a stronger reaction
by regulators to climate change. But there is still the question why has this
not happened much earlier and how long will it take until there are
significant effects.

Our hypothesis is that the evolution of social norms is a slow process, and
their transmission between different social groups is also complicated. We
start from a situation in which, as \cite{carney2015breaking} pointed out \textquotedblleft The
horizon for monetary policy extends out to 2-3 years. For financial
stability it is a bit longer, but typically only to the outer boundaries of
the credit cycle -- about a decade.\textquotedblright\ If that is the status
quo (social norm) about appropriate actions by central banks, it is difficult to expect the regulators to start taking a view that
goes perhaps to half a century or more.

But even if norms are slow in changing, they do change. A recent study shows
that women are now seen as equal or more competent than men, something that
didn't happen half a century ago. A similar thing happens with same-sex
marriage. These changes in attitudes are now encoded in regulations
fostering gender equality in corporate boards, or laws allowing same-gender
marriage. But it gets even better. For environmental protection both
farmers, and businesses in general, often go beyond legal mandates. And as
\cite{gunningham2004social} say: \textquotedblleft the increasing
incidence of \textquotedblleft beyond compliance\textquotedblright\
corporate behavior can be better explained in terms of the interplay between
social pressures and economic constraints.\textquotedblright\ 

Our project approach to answering the question for how norms change and
diffuse between groups starts by proposing a model of norms transmission in
social networks. We assume that individuals take actions that have an
(idiosyncratic) benefit and a cost. In addition, there is a complementarity
between the actions of the individual and those of others in her group and
in other groups that are \textquotedblleft close\textquotedblright\ to them
or whose opinions are important. The model has a simple linear quadratic
structure (as in \cite{ballester2006s}) and delivers
a unique equilibrium where the actions of group members depend on their
idiosyncratic preferences and those of others in close groups. Given its
structure, the model's parameters can be easily identified through an
econometric model. 

We complement the analytical framework for the problem with its
empirical analysis. The aim of this part of the project is to ascertain the
web of influences between different actors in climate change policy. We have
collected information (using advanced web-scraping methods) about mentions
to climate change in mainstream news media (from the US, UK, Germany, France, and
Spain), general interest scientific journals (Nature, Science), top Economics journals,\footnote{
The so-called top 5: \emph{Quarterly Journal of Economics}, \emph{American Economic
Review}, \emph{Journal of Political Economy}, \emph{Econometrica} and \emph{Review of Economic
Studies}.} European Parliament questions, and European Central Bank
presidential speeches, since the 1990s. We then build a Vector Auto
Regressive model (VAR) to estimate how the mentions in one of these actors
in one period are correlated with lagged mentions by other actors.

In terms of descriptive evidence, we have found that natural scientists had
been concerned with the problem since more than 30 years ago, academic
economists are generally unconcerned even now, the mainstream media and the European Parliament started worrying seriously about the problem about the turn of the century, and the ECB increased their concern in very recent years.

In terms of the analytical results from the VAR, we study the data at
quarterly frequency. Three of our variables are mentions about climate change in
different outlets: the news media, Euro parliament, and general interest scientific journals. We also use GDP as a control variable.
We find that media and the parliament are mutually affected. Other than that, we also find strong interactions with GDP fluctuations. This is a concern. A long term problem like climate change should not ebb and flow with relatively small (in the grand scheme of things) output fluctuations. But the finding can be a tool for concerned organizations to focus the resources at times of social inattention. 

We cannot find influences of science on media or parliament. It is tempting to think scientific efforts are useless in this domain, particularly given the slow motion of regulatory responses. But we need to be cautious, it could also be that the influences are more subtle than the statistical model can capture.

\subsection{Related literature}

This paper contribute to several strands of the literature. One of them is the one related to social norms. 
\cite{fehr2018normative} have argued that many regularities regarding cooperation can be explained if individuals hold a social norm of conditional cooperation (\cite{kimbrough2016norms} and \cite{kolle2020promoting}, \cite{szekely2021evidence} provide evidence of norm-following that leads to cooperation). In fact, social norms have been proposed as a key instrument to solve social dilemmas (\cite{ostrom2000collective}; \cite{bicchieri2005grammar}; \cite{biel2007activation}) in general, and  climate change in particular \cite{riehm2020social}. We contribute to this literature by providing a model and evidence showing how those norms spread in the population.

We also contribute to a large literature about the media communication of climate change (\cite{wilson2013communicating}, \cite{gavin2009addressing}). To this literature we provide a comprehensive view of the evolution of the coverage and its interaction with other domains. A similar contribution is provided to the literature on scientific journals coverage of climate change (including the surprisingly low coverage in top economics journals) as in \cite{nielsen2011news}, \cite{ladle2005scientists}, \cite{oswald2019does}, or in political circles \cite{willis2017taming}, \cite{willis2018constructing}, and central banks \cite{olovsson2018climate}, \cite{skinner2021central}.

Our method for creating indices is taken from \cite{bloom} and \cite{ghirelli2021measuring} applied to a different field. Our theoretical model is inspired by the work in social networks pioneered by \cite{ballester2006s}

\section{Evolution of mentions to climate change}

In this section we provide a visual description of the evolution of climate change
mentions in different sectors: the news media, the Euro parliament, scientific journals,
and ECB speeches. This is our proxy for the preoccupation about climate change in those
sectors.

\subsection{Developing a Climate Change Index of Public Interest}

We analyze the presence of Climate Change and their evolution over time for the main American and European newspapers. \cite{bloom} manages to measure an unobservable variable, such as uncertainty in Economic Policy, with an idea as simple as it is powerful: the level of impact that this variable has is reflected in the repetition of terms related to economic uncertainty in the different newspapers over time. The more these terms are used, the more impact/interest the variable is having in that period. In a similar way, we develop a Climate Change Index (CCI) using the universe of news in top European and United States newspapers using the keywords "climate change". It is trivial to observe, due to the nature of these words, that any text that uses them will be alluding to this problem, making identification very simple.

Following \cite{bloom}, we standardize the monthly shares newspaper-level series to unit standard deviation from 1995 to 2021 and then average across the 12 European papers by month. Finally, we divide this average by the mean and multiply by 100 for the same period to obtain the normalized series. 

\begin{figure}[h!]
\centering
\includegraphics[scale=0.95]{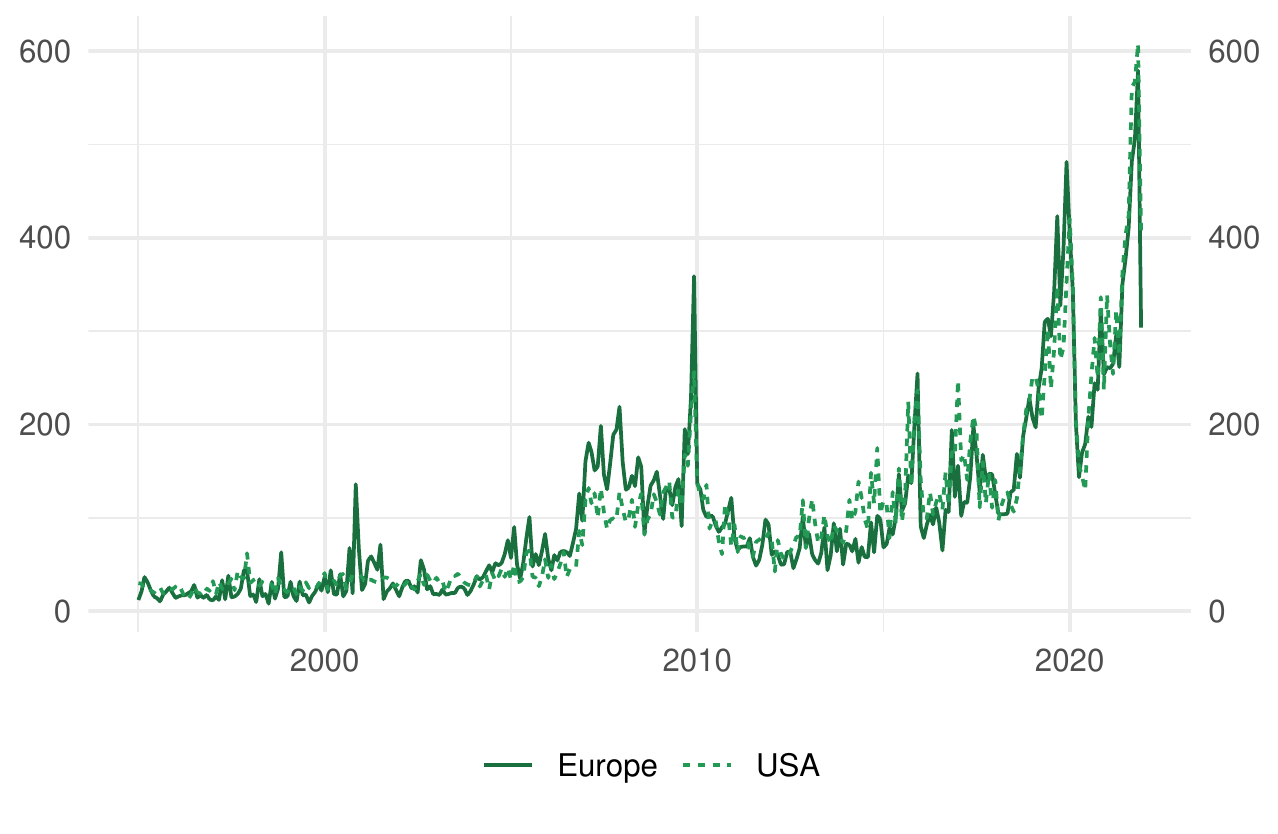}
\caption{Monthly Climate Change Index for Europe and the USA}
\label{fig:media}
\end{figure}

Figure \ref{fig:media} shows the Climate Change Index for Europe and the US, which have a correlation of 0.947. As we can see media does not show interest until around 2003, but it is not until 2015 that it becomes part of the relevant topics. It is interesting to notice how the media seems to respond to trends; there are peaks in which it pays attention to climate change, and other periods in which the intensity drops even though the problem has not been solved, but on the contrary seems that the problem has become worse.

In the appendix we include the shares that climate change news occupy for the different newspapers that make up the index, in addition to an index for each country. 

\subsection{Climate Change and Top 5 journals in Economics}

We count the number of papers published in Top 5 journals in Economics that use "Global Warming" or "Climate Change" in their abstract for the period 1999-2021. The results speak for themselves about economists' interest in the topic. Since the data is about published papers, it is difficult to know to which extent referees and editors are responsible for this, as we do not have data on submissions. It could be that climate change related papers have a higher proportion relative to total submissions.

\begin{sidewaystable} 

\centering
\caption{Count of words in the Top-5 Journals in Economics}
\label{tab:2}
\begin{tabular}{lccccccccccccccccccccccc}
  \hline
  \hline
 & 99 & 00 & 01 & 02 & 03 & 04 & 05 & 06 & 07 & 08 & 09 & 10 & 11 & 12 & 13 & 14 & 15 & 16 & 17 & 18 & 19 & 20 & 21 \\   \hline
  Climate Change & 0 & 0 & 0 & 0 & 0 & 0 & 0 & 0 & 1 & 0 & 0 & 0 & 1 & 2 & 0 & 1 & 2 & 2 & 0 & 0 & 1 & 1 & 0 \\ 
  Global Warming & 0 & 0 & 0 & 0 & 0 & 0 & 0 & 0 & 0 & 0 & 0 & 0 & 0 & 0 & 0 & 0 & 0 & 1 & 0 & 0 & 0 & 0 & 0 \\ 
  Systemic Risk & 0 & 0 & 0 & 0 & 0 & 0 & 0 & 0 & 0 & 1 & 0 & 0 & 0 & 0 & 0 & 0 & 1 & 1 & 0 & 0 & 1 & 0 & 0 \\ 
  Environmental & 2 & 3 & 3 & 0 & 0 & 4 & 1 & 3 & 1 & 2 & 2 & 1 & 2 & 7 & 2 & 2 & 2 & 3 & 1 & 4 & 3 & 4 & 5 \\ 
  Pollution & 0 & 0 & 1 & 1 & 1 & 0 & 2 & 1 & 0 & 2 & 3 & 1 & 2 & 2 & 1 & 1 & 1 & 4 & 2 & 4 & 2 & 2 & 3 \\ 
  Carbon Tax & 0 & 0 & 0 & 0 & 0 & 0 & 0 & 0 & 0 & 0 & 0 & 0 & 0 & 1 & 0 & 1 & 0 & 2 & 1 & 1 & 1 & 1 & 0 \\ 
  Optimal Taxation & 0 & 0 & 1 & 0 & 0 & 0 & 0 & 1 & 2 & 1 & 1 & 2 & 0 & 0 & 0 & 0 & 0 & 3 & 0 & 0 & 2 & 2 & 0 \\ 
  Countercyclical & 2 & 2 & 0 & 1 & 1 & 4 & 2 & 1 & 3 & 0 & 2 & 1 & 0 & 8 & 2 & 3 & 1 & 2 & 5 & 5 & 1 & 2 & 1 \\ 
  Gold Standard & 0 & 0 & 1 & 0 & 1 & 0 & 1 & 0 & 0 & 1 & 0 & 0 & 0 & 0 & 0 & 0 & 0 & 0 & 1 & 0 & 0 & 0 & 0 \\ 
  Corruption & 1 & 1 & 0 & 2 & 2 & 0 & 2 & 1 & 3 & 1 & 1 & 0 & 1 & 0 & 1 & 2 & 3 & 2 & 0 & 2 & 0 & 2 & 2 \\ 
  Unemployment &  3 &  2 &  5 &  2 &  4 &  7 &  5 &  5 &  8 &  7 &  6 &  9 &  7 &  5 &  7 &  6 &  7 & 11 & 13 & 12 &  6 &  9 &  7 \\ 
  Marketing & 0 & 0 & 2 & 0 & 2 & 1 & 0 & 2 & 0 & 0 & 1 & 2 & 2 & 3 & 0 & 1 & 0 & 1 & 1 & 1 & 1 & 1 & 1 \\ 
  Monetary Policy &  2 &  4 &  3 &  5 & 10 &  4 & 10 &  4 &  3 &  4 &  4 &  5 &  6 &  7 &  4 &  3 &  3 & 10 &  7 &  8 &  8 & 15 &  7 \\ 
  Game Theory & 1 & 2 & 0 & 3 & 2 & 1 & 1 & 1 & 3 & 0 & 0 & 1 & 2 & 1 & 1 & 0 & 1 & 2 & 0 & 0 & 0 & 0 & 1 \\ 
  Optimal Policy & 3 & 1 & 1 & 1 & 1 & 0 & 1 & 1 & 1 & 4 & 2 & 4 & 2 & 5 & 2 & 1 & 1 & 4 & 0 & 2 & 3 & 2 & 1 \\ 
  Inflation &  1 &  6 &  8 &  7 & 12 &  3 & 12 &  7 &  7 &  7 &  9 &  2 & 10 &  3 &  3 &  2 &  5 &  4 &  7 &  8 & 12 &  6 &  7 \\ 
  Tax  &  9 &  4 &  7 &  9 & 10 &  4 & 13 & 10 &  5 & 11 & 13 &  8 &  9 & 11 & 21 & 11 & 16 & 20 & 19 & 14 & 28 & 16 & 21 \\ 
  Inequality & 10 & 11 &  6 & 11 & 10 & 12 &  7 & 10 &  5 &  5 &  9 & 10 &  7 &  4 & 11 &  7 &  8 & 14 & 15 & 15 & 10 & 17 &  9 \\ 
  Transportation & 0 & 0 & 1 & 0 & 0 & 2 & 1 & 0 & 1 & 1 & 2 & 0 & 2 & 1 & 0 & 0 & 0 & 1 & 0 & 2 & 2 & 1 & 0 \\ 
  Institutions &  5 &  3 &  2 &  7 &  5 &  3 &  7 &  6 &  6 &  7 &  7 &  7 & 10 &  8 &  7 &  8 &  7 &  8 &  7 &  3 &  9 &  6 &  3 \\ 
  WWII & 0 & 0 & 0 & 1 & 1 & 0 & 0 & 1 & 0 & 1 & 0 & 0 & 1 & 0 & 1 & 0 & 0 & 1 & 0 & 0 & 0 & 1 & 0 \\ 
   \hline
\end{tabular}

\end{sidewaystable} 

\newpage
\subsection{Climate Change and General Interest Scientific Journals}

Following the methodology of the CCI for media, we construct an index using the main General Interest Scientific Journals, Nature and Science between 1995-2021. As we can see, interest in climate change is clearly growing with an stable trend over time, unlike what happens with the media, which seems to respond to behavioral criteria.

\begin{figure}[h!]
\centering
\includegraphics[scale=0.95]{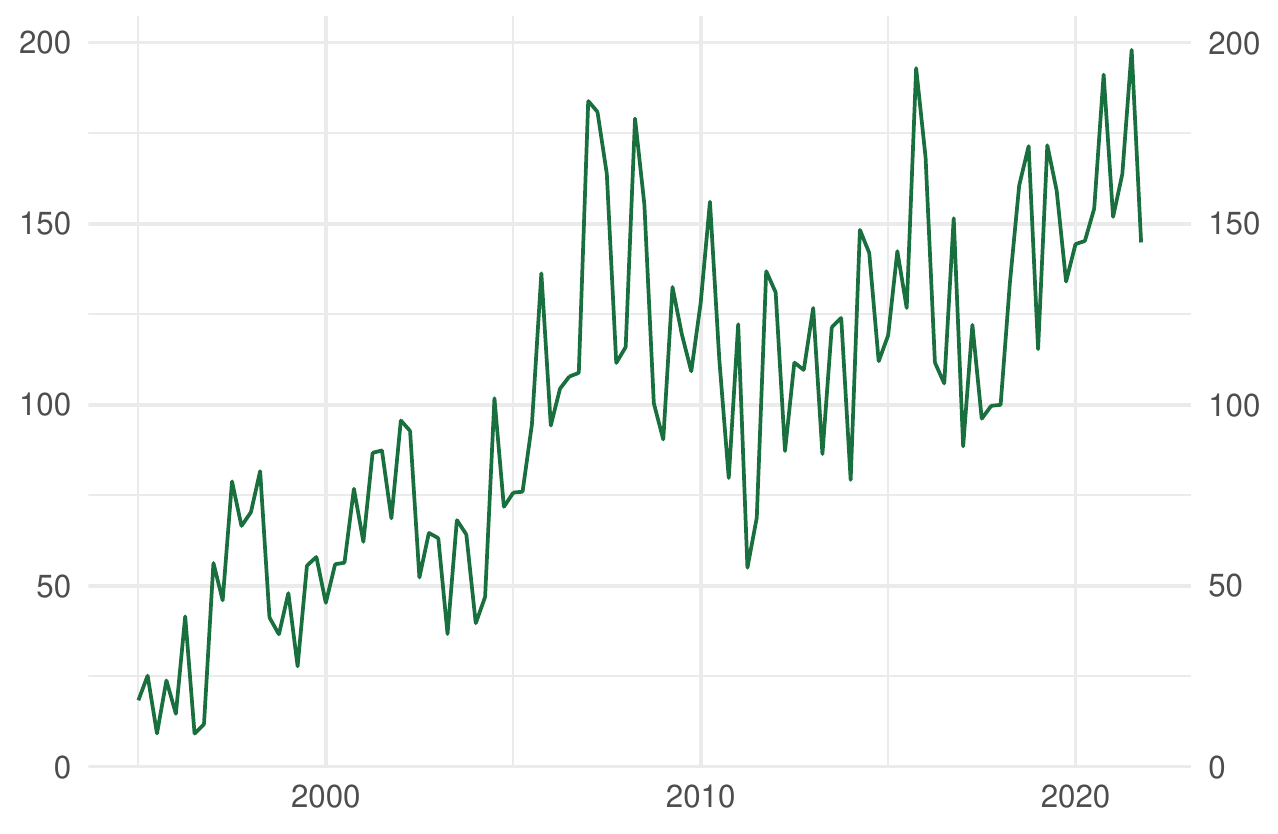}
\caption{Quarterly Climate Change Index for the General Interest Science Journals.}
\label{fig:scienceandnature}
\end{figure}

\newpage
\subsection{European Parliament}

We create a new source of information from the share of questions made in the European Parliament for the period 1995-2021 containing the words "climate change", normalized to have standard deviation and mean 100.

\begin{figure}[h!]
\centering
\includegraphics[scale=0.95]{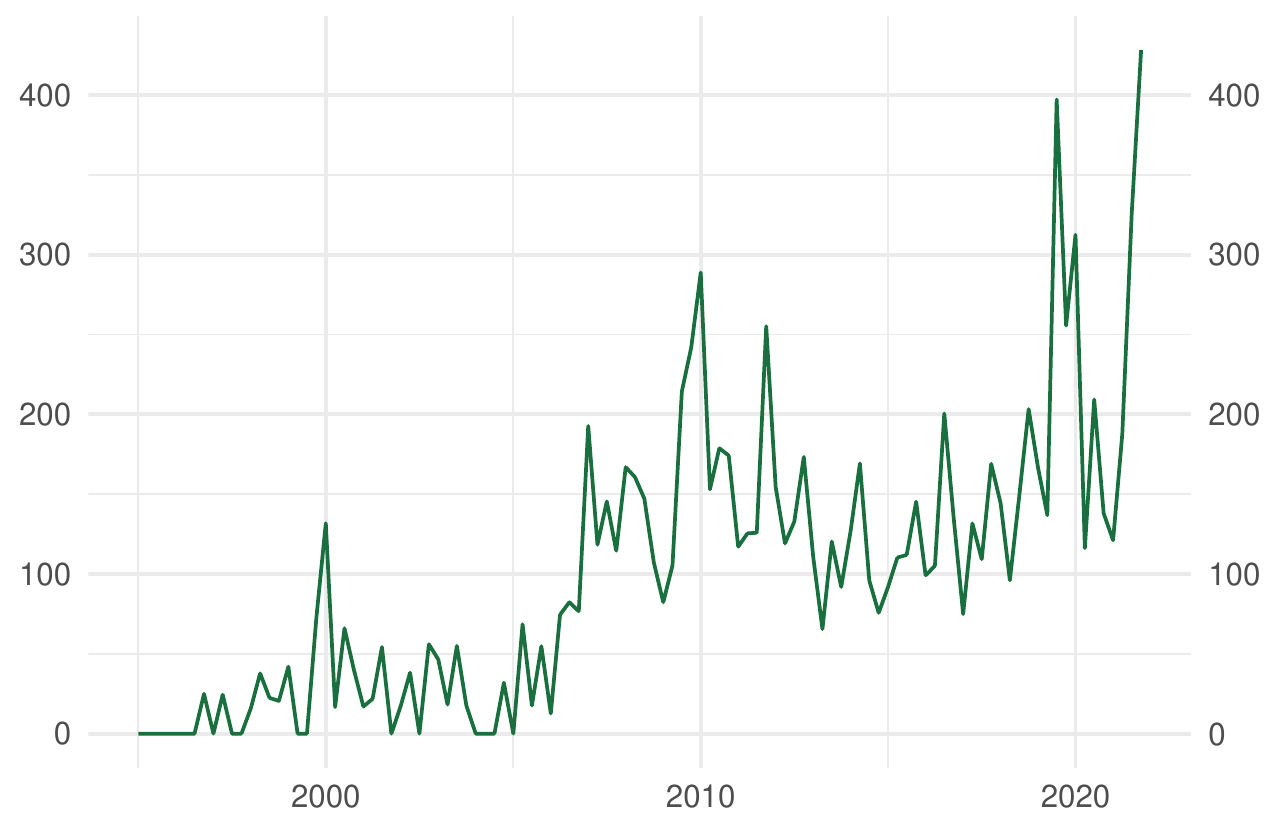}
\caption{Quarterly Climate Change Index for the European Parliament}
\label{fig:parliament}
\end{figure}

\subsection{Central Bank Speeches}

\begin{figure}[h!]
\centering
\includegraphics[scale=0.95]{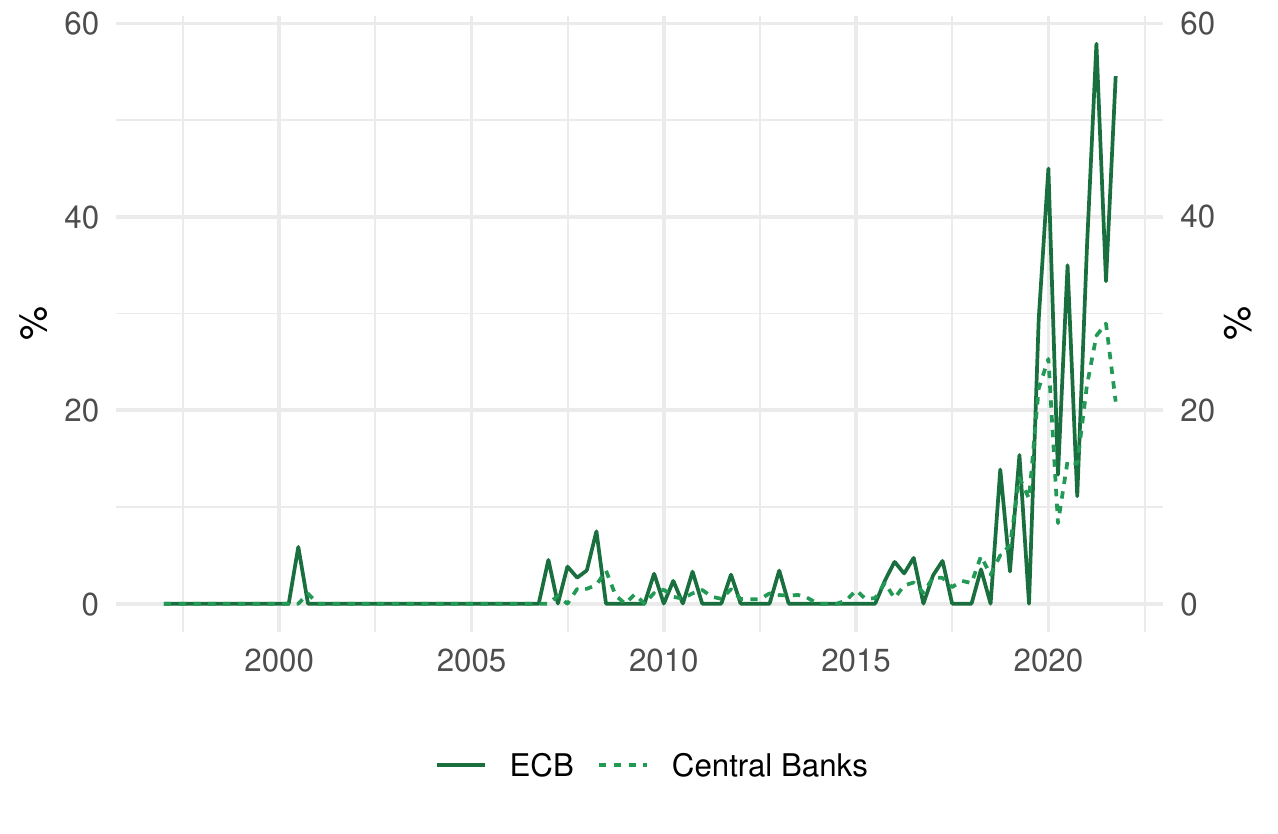}
\caption{Share of Speeches containing the words "Climate Change"}
\label{fig:ECBspeeches}
\end{figure}

We count the share of ECB presidential speeches\footnote{Available at \href{https://www.ecb.europa.eu/press/key/html/downloads.en.html}{ECB site}.} in English mentioning the words "Climate Change" for the period since its creation in 1997, also, from the Central bankers' speeches available at the BIS repository\footnote{The  \href{https://www.bis.org/cbspeeches/index.htm}{BIS} site contains more then 17,000 speeches in English from the Federal Reserve, ECB, and many Central Banks.}.

Until 2018, these keywords were practically not used, demonstrating the lack of interest in the subject. However, since 2019, more than 50\% of the ECB speeches and around a 25\% of the total Central Bankers speeches already included these words, showing that although it cannot be analyzed in this paper due to the scarcity of data, climate change has become a crucial issue for the Central Banks.

In the Appendix we show the results of comparing mentions of climate change in ECB speeches with mentions of other relevant terms, like ``taxes'' or ``inequality''. Taxation is mentioned very frequently from the beginning, inequality is less frequent, but it starts earlier than climate change. Strikingly, climate change is now more frequently mentioned than either taxes or inequality.

\subsection{The Federal Open Market Committee - FOMC}

As a counterpoint to the ECB speeches we count the number of questions made in the transcripts from the FOMC for "Climate Change" for the period 1975-2015 (the transcripts are available only 5 years after). "Climate Change" appear only once (related to climate) \href{https://www.federalreserve.gov/monetarypolicy/files/FOMC20091216meeting.pdf}{here}. Nevertheless, in the minutes published by the FOMC from 1993 to present, it only appear twice, in \href{https://www.federalreserve.gov/monetarypolicy/fomcminutes20191030.htm}{2019} and \href{https://www.federalreserve.gov/monetarypolicy/fomcminutes20201105.htm}{2020}.

\section{A simple theoretical framework}\label{theory}

In order to understand the relationship between the different institutions and social group whose preoccupation with climate we characterize with their public utterances, we first describe a tractable model which we later estimate using a vector auto-regression (VAR).

Every individual $j$ belongs to some group $G_{j}$ where $|G_{j}|\in R$. A parameter 
$\lambda _{G_{i}G_{j}}$ measures how a group $i$ person cares about a group $j$ person.
Every individual experiences an idiosyncratic amount of intrinsic interest in the policy $b_{i}$.
There is a costly action $a_{i_{t}}$ that each individual takes in every period $t$. This action has a cost per unit $c_{i}$.
With these elements in place, we can write the utility function as:
\[
U\,_{i}\left( a_{i_{t}},\mathbf{a}_{t-1}\right) =a_{i_{t}}\left(
b_{i}+\sum_{j\in R}\lambda _{G_{i}G_{j}}a_{j_{t-1}}\right) -\frac{c_{i}}{2}%
a_{i_{t}}^{2}
\]%
Then, the optimal action for each individual can be written as:
\[
a_{it}=\frac{1}{c_{i}}\left( b_{i}+\sum_{j\in R}\lambda
_{G_{i}G_{j}}a_{j_{t-1}}\right) 
\]%
And since the individual actions are linear in others' previous actions, we can aggregate to an institutional level (a key assumption in this case is that the interaction parameters $\lambda _{G_{i}G_{j}}$ are common within groups).
Given this, the VAR constant in the equation for each group's "action" (the number
of messages) is $b_{i}/c_{i}$, i.e. the intrinsic interest in the policy
(relative to the cost of messaging) and a coefficient of the action of
other groups is $\lambda _{G_{i}G_{j}}/c_{i}$ i.e. the impact on the
marginal benefit of group $G_{i}$ of an increase in $G_{j}$ action (relative
to the cost). We have introduced just one lag in this description but, of course, we can write as
many as we want. Also we have written lagged actions in the utility
function, but we can also write expectations and say that the expectations
are formed naively so that 
\[
E\left( \mathbf{a}_{t}\right) =\mathbf{a}_{t-1}.
\]

\section{VAR model estimation}

To understand the interconnection between the different actors we estimate a VAR micro-founded from the model in Section \ref{theory}. It can be written as $X_t = \Pi(L)X_t + \epsilon_t$, where $X_t$ is a set of endogenous variables,  $\Pi$ is a matrix of VAR coefficients capturing the dynamics of the system, and $\epsilon_t : N(0, \Sigma)$ is a vector of shocks having zero mean and variance–covariance matrix $\Sigma$. The variables in $X_t$ are
the following:  $x_1$ is mentions of climate change in the media (CCI), $x_2$  is mentions in the European parliament questions (normalized), $x_3$ is mentions in science journals (CCI), and $x_4$ is GDP for the Euro Area (normalized).

Table \ref{tab:VAR} displays the results. The notation ARx(y,z) means that "x" is the lag, "y" the index of the variable whose effect we measure, and z is the index of the variable affected by it.

The data is quarterly, and at one quarter all variables are affected by their own lags. The Euro Parliament positively affects the media. This means that an increase in the debate in the European Parliament on climate change translates into an increase in the media interest in the following quarter.

At two quarters there are no own effects. We also find a negative effect of Euro Parliament on media, which can be interpreted as the loss of media interest on climate change after a quarter. There is also a reciprocal negative effect of media on Euro parliament.

At three quarters the only own effects are given by scientific journals and GDP. There is a negative effect of GDP in the Euro Parliament. Six months after a boom they forget about climate change, but they mention it more during a recession.

At four quarters the only own effect is given by GDP, and there is only one positive effect from GDP to the Euro parliament. This can be interpreted as the lack of persistence of the effect appeared at three quarters.

Generally speaking, we find that media is affected by the parliament, and parliament is affected by the media. Other than that, we also find very strong interactions with GDP fluctuations. This is worrying since attention in a long term issue like climate change should not be driven by short time fluctuations in economic activity. But it is an important finding as it suggests a time when activists should concentrate their efforts. Science, on the other hand, seems to have no discernible effect on either parliament or the media. This is probably because the influences of Science are more subtle and long-term than the statistical model can uncover.

\begin{table}
\centering
\caption{VAR: ARx(y,z) "x" is the lag, "y" is affecting variable, and "z" is affected variable\\}
\label{tab:VAR}
\begin{tabular}{lcccc}
  \hline
  \hline
&Value&Standard Error&TStatistic&PValue\\
  \hline

Constant(1)&-101.51&56.43&-1.8&0.07\\
Constant(2)&-92.19&78.69&-1.17&0.24\\
Constant(3)&-93.88*&41.64&-2.25&0.02\\
Constant(4)&1.54&2.59&0.6&0.55\\
AR{1}(1,1)&0.82***&0.12&6.91&0\\
AR{1}(2,1)&0.61***&0.17&3.69&0\\
AR{1}(3,1)&0.01&0.09&0.15&0.88\\
AR{1}(4,1)&0&0.01&-0.39&0.7\\
AR{1}(1,2)&-0.07&0.08&-0.88&0.38\\
AR{1}(2,2)&0.24**&0.11&2.08&0.04\\
AR{1}(3,2)&0&0.06&0.01&0.99\\
AR{1}(4,2)&-0.01&0&-1.55&0.12\\
AR{1}(1,3)&0.06&0.14&0.42&0.68\\
AR{1}(2,3)&-0.17&0.19&-0.88&0.38\\
AR{1}(3,3)&0.39***&0.1&3.88&0\\
AR{1}(4,3)&0.01&0.01&1.48&0.14\\
AR{1}(1,4)&3.23&2.23&1.45&0.15\\
AR{1}(2,4)&0.3&3.11&0.1&0.92\\
AR{1}(3,4)&0.91&1.65&0.55&0.58\\
AR{1}(4,4)&0.79***&0.1&7.76&0\\
AR{2}(1,1)&0.23&0.15&1.57&0.12\\
AR{2}(2,1)&-0.44*&0.21&-2.1&0.04\\
AR{2}(3,1)&0.11&0.11&0.96&0.34\\
AR{2}(4,1)&0&0.01&-0.57&0.57\\
AR{2}(1,2)&-0.25&0.08&-2.97&0\\
AR{2}(2,2)&0.16&0.12&1.35&0.18\\
AR{2}(3,2)&-0.02&0.06&-0.38&0.71\\
AR{2}(4,2)&0&0&-0.38&0.71\\
   \hline
\end{tabular}

\end{table}

\begin{table}
\centering
\caption{VAR: ARx(y,z) "x" is the lag, "y" is affecting variable, and "z" is affected variable\\}
\label{tab:VAR2}
\begin{tabular}{lcccc}
  \hline
  \hline
&Value&Standard Error&TStatistic&PValue\\
  \hline
AR{2}(1,3)&0.2&0.14&1.43&0.15\\
AR{2}(2,3)&-0.1&0.2&-0.48&0.63\\
AR{2}(3,3)&-0.12&0.11&-1.12&0.26\\
AR{2}(4,3)&0.01&0.01&1.28&0.2\\
AR{2}(1,4)&-4.44&2.55&-1.74&0.08\\
AR{2}(2,4)&1.21&3.56&0.34&0.73\\
AR{2}(3,4)&-2.39&1.88&-1.27&0.2\\
AR{2}(4,4)&0.02&0.12&0.2&0.84\\
AR{3}(1,1)&-0.11&0.15&-0.72&0.47\\
AR{3}(2,1)&0.22&0.21&1.08&0.28\\
AR{3}(3,1)&-0.15&0.11&-1.32&0.19\\
AR{3}(4,1)&0&0.01&0.45&0.65\\
AR{3}(1,2)&-0.1&0.09&-1.21&0.23\\
AR{3}(2,2)&-0.07&0.12&-0.62&0.54\\
AR{3}(3,2)&0.01&0.06&0.15&0.88\\
AR{3}(4,2)&-0.02***&0&-4.2&0\\
AR{3}(1,3)&-0.14&0.14&-1&0.32\\
AR{3}(2,3)&0.26&0.2&1.3&0.19\\
AR{3}(3,3)&0.26*&0.11&2.46&0.01\\
AR{3}(4,3)&0&0.01&-0.39&0.7\\
AR{3}(1,4)&3.37&2.52&1.34&0.18\\
AR{3}(2,4)&3.45&3.51&0.98&0.33\\
AR{3}(3,4)&2.77&1.86&1.49&0.14\\
AR{3}(4,4)&0.38**&0.12&3.27&0\\
AR{4}(1,1)&0.12&0.13&0.97&0.33\\
AR{4}(2,1)&-0.04&0.18&-0.22&0.83\\
AR{4}(3,1)&0.09&0.1&0.94&0.35\\
AR{4}(4,1)&0&0.01&0.43&0.67\\
   \hline
\end{tabular}

\end{table}

\begin{table}
\centering
\caption{VAR: ARx(y,z) "x" is the lag, "y" is affecting variable, and "z" is affected variable\\}
\label{tab:VAR3}
\begin{tabular}{lcccc}
  \hline
  \hline
&Value&Standard Error&TStatistic&PValue\\
  \hline
AR{4}(1,2)&0.15&0.09&1.79&0.07\\
AR{4}(2,2)&0.13&0.12&1.06&0.29\\
AR{4}(3,2)&-0.04&0.06&-0.58&0.56\\
AR{4}(4,2)&0.02***&0&4.68&0\\
AR{4}(1,3)&0&0.14&-0.03&0.98\\
AR{4}(2,3)&0.06&0.2&0.28&0.78\\
AR{4}(3,3)&-0.04&0.1&-0.43&0.67\\
AR{4}(4,3)&-0.01&0.01&-0.87&0.39\\
AR{4}(1,4)&-1.03&2.17&-0.47&0.64\\
AR{4}(2,4)&-3.85&3.03&-1.27&0.2\\
AR{4}(3,4)&0.19&1.6&0.12&0.91\\
AR{4}(4,4)&-0.21*&0.1&-2.1&0.04\\

   \hline
\end{tabular}

\end{table}

\newpage
\section{Conclusion}
We have documented the evolution of mentions of climate change in different environments: policy, sciences, and the general public (proxied by news media). We have also postulated a model about how those different environments influence one another and then estimated the model's parameters. We find large fluctuations of interest and interesting cross influences. A particularly salient one is related to how GDP evolution affects the interest in climate change. These observations could be a useful tool for timing activists and other groups interested in influencing social debate.

Future research could expand our results by doing a more fine grained analysis of the connections inside the different groups, potentially using tools from social complex network analysis.

\newpage

\bibliography{references}

\newpage
\section*{Appendix A. A CCI for each country.}

\paragraph{United Kingdom}

We use the keywords "Climate Change" for The Guardian, The Times, The Sun, and The Independent.

\begin{figure}[h!]
\centering
\includegraphics[scale=0.95]{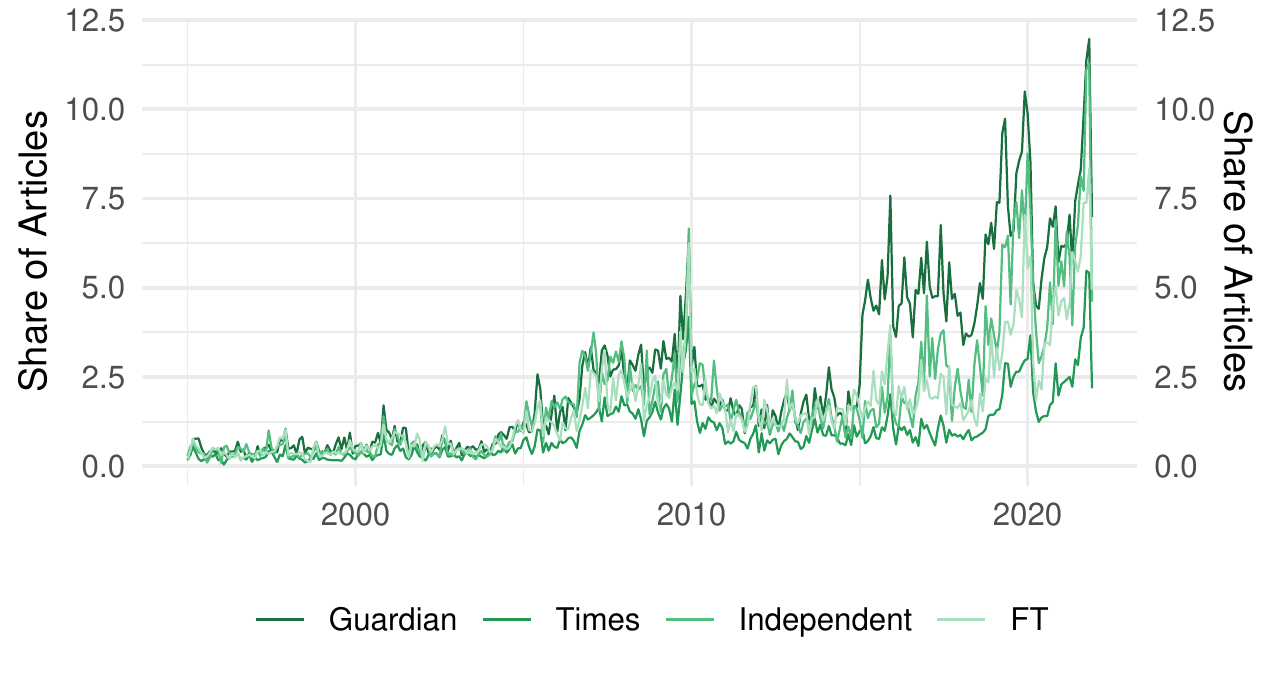}
\caption{Share of Articles in the British Media}
\label{fig:uk_med}
\end{figure}

\paragraph{Spain}

We use the keywords "Cambio Climático (Climate Change)" for El Mundo, El País, and ABC.

\begin{figure}[h!]
\centering
\includegraphics[scale=0.95]{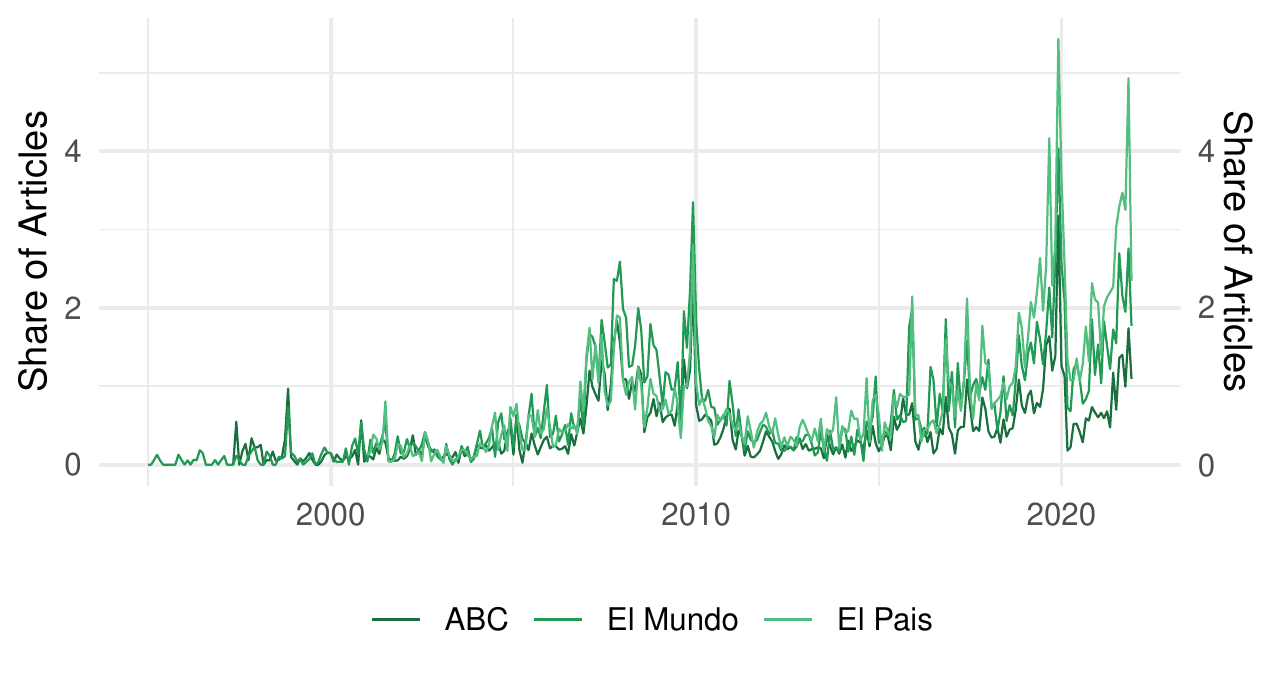}
\caption{Share of Articles in the Spanish Media}
\label{fig:uk_med}
\end{figure}

\newpage

\paragraph{Germany}

We use the keywords "Klimawandel (Climate Change)" and how much they have been used in Bild and DIE ZEIT.

\begin{figure}[h!]
\centering
\includegraphics[scale=0.95]{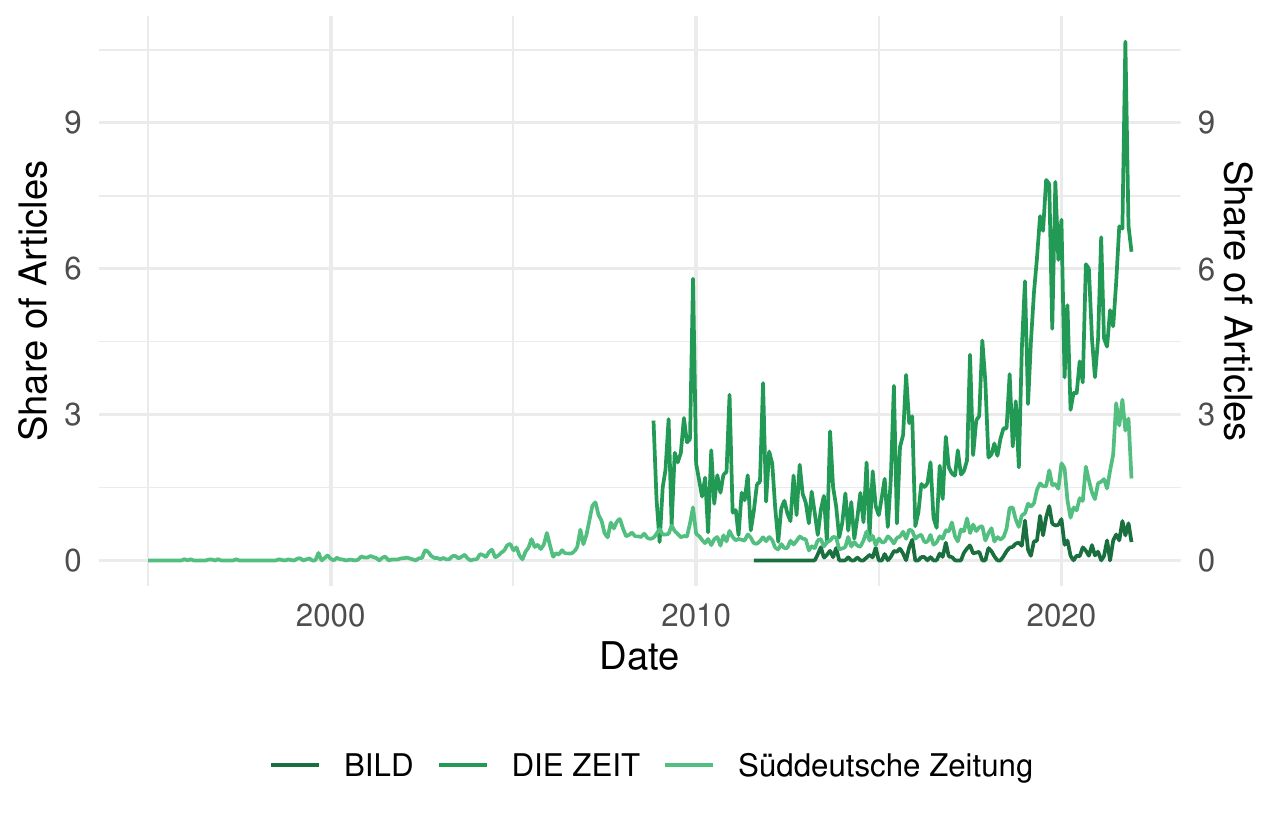}
\caption{Share of Articles in the German Media}
\label{fig:uk_med}
\end{figure}

\paragraph{France}

We use the keywords "Changement Climatique (Climate Change)" and how much they have been used in Le Figaro and Les Echos

\begin{figure}[h!]
\centering
\includegraphics[scale=0.95]{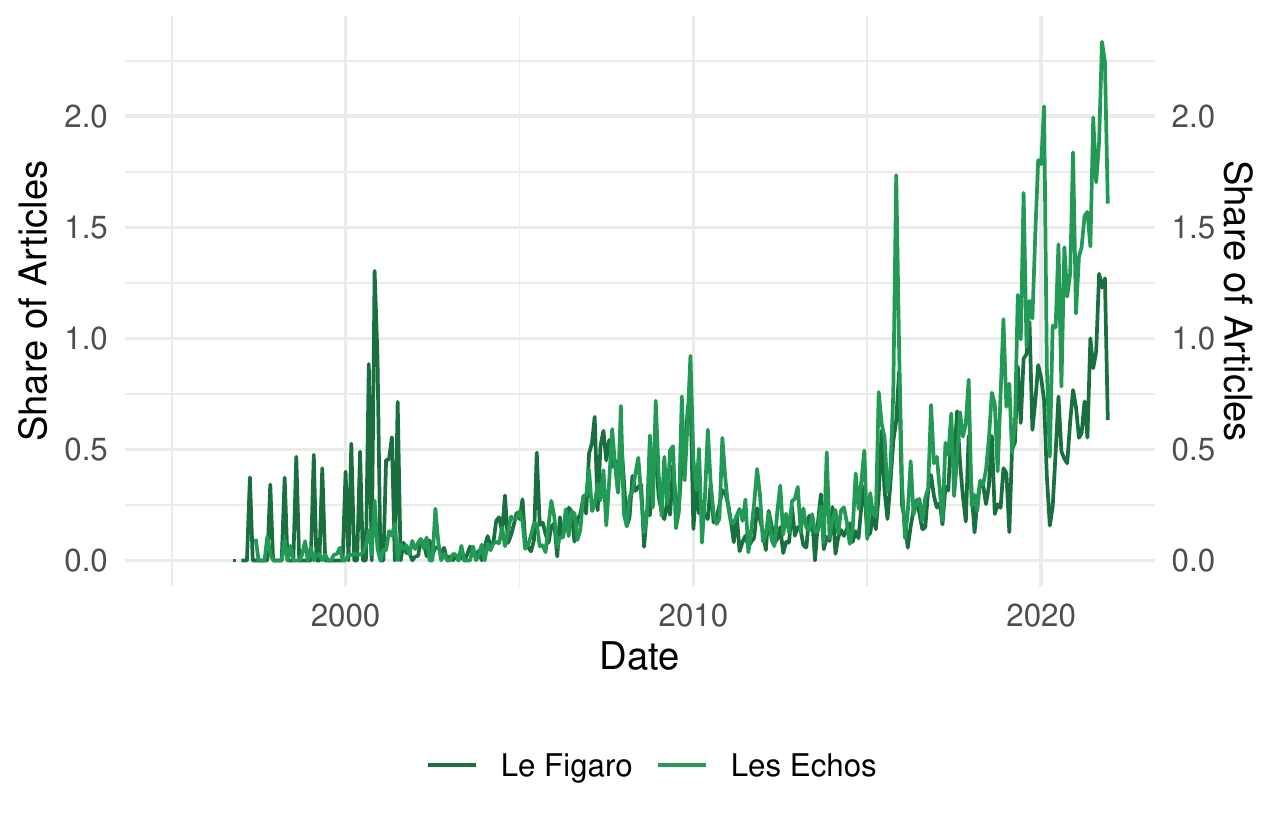}
\caption{Share of Articles in the French Media}
\label{fig:uk_med}
\end{figure}

\paragraph{Italy}

We use the keywords "Cambiamento Climatico (Climate Change)" and how much they have been used in Corriere della Sera, and La Repubblica.

\begin{figure}[h!]
\centering
\includegraphics[scale=0.95]{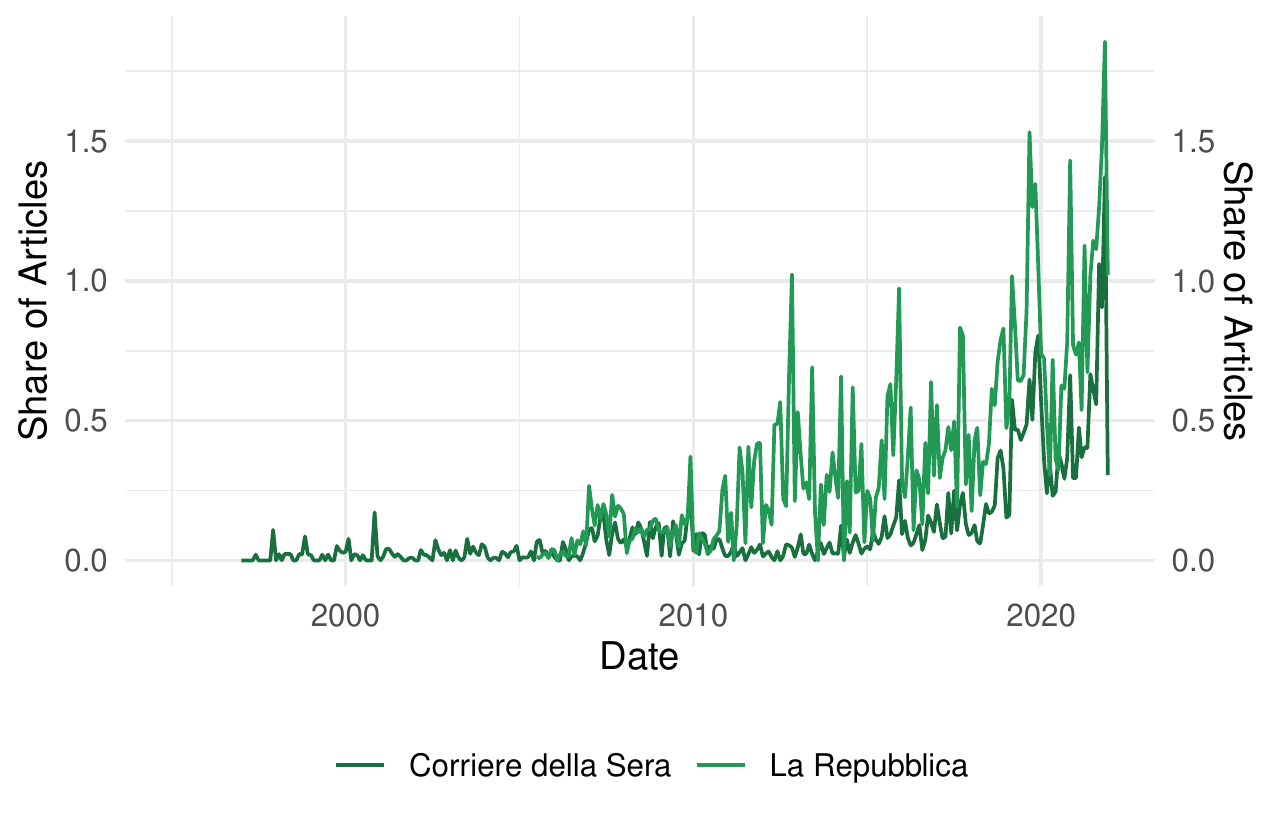}
\caption{Share of Articles in the Italian Media}
\label{fig:uk_med}
\end{figure}

\begin{figure}[h!]
\centering
\includegraphics[scale=0.85]{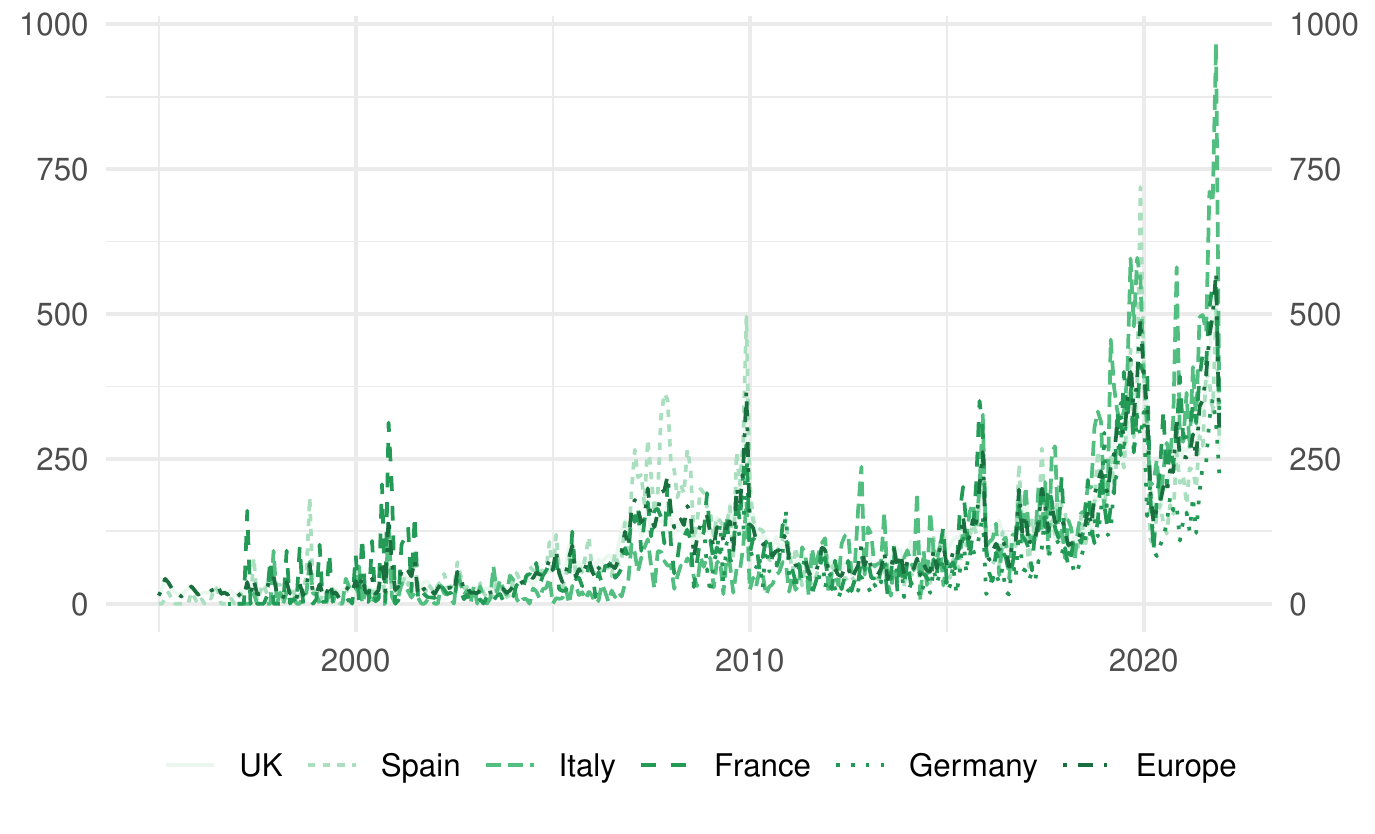}
\caption{Climate Change Index for the European Countries}
\label{fig:uk_med}
\end{figure}

\newpage
\section*{Appendix B. Natural Science Journals.}

\paragraph{Science}

We count the number of articles (total) published in Science and Nature that use "Climate Change" in their abstract for the period 1995-2021.

\begin{figure}[h!]
\centering
\includegraphics[scale=0.95]{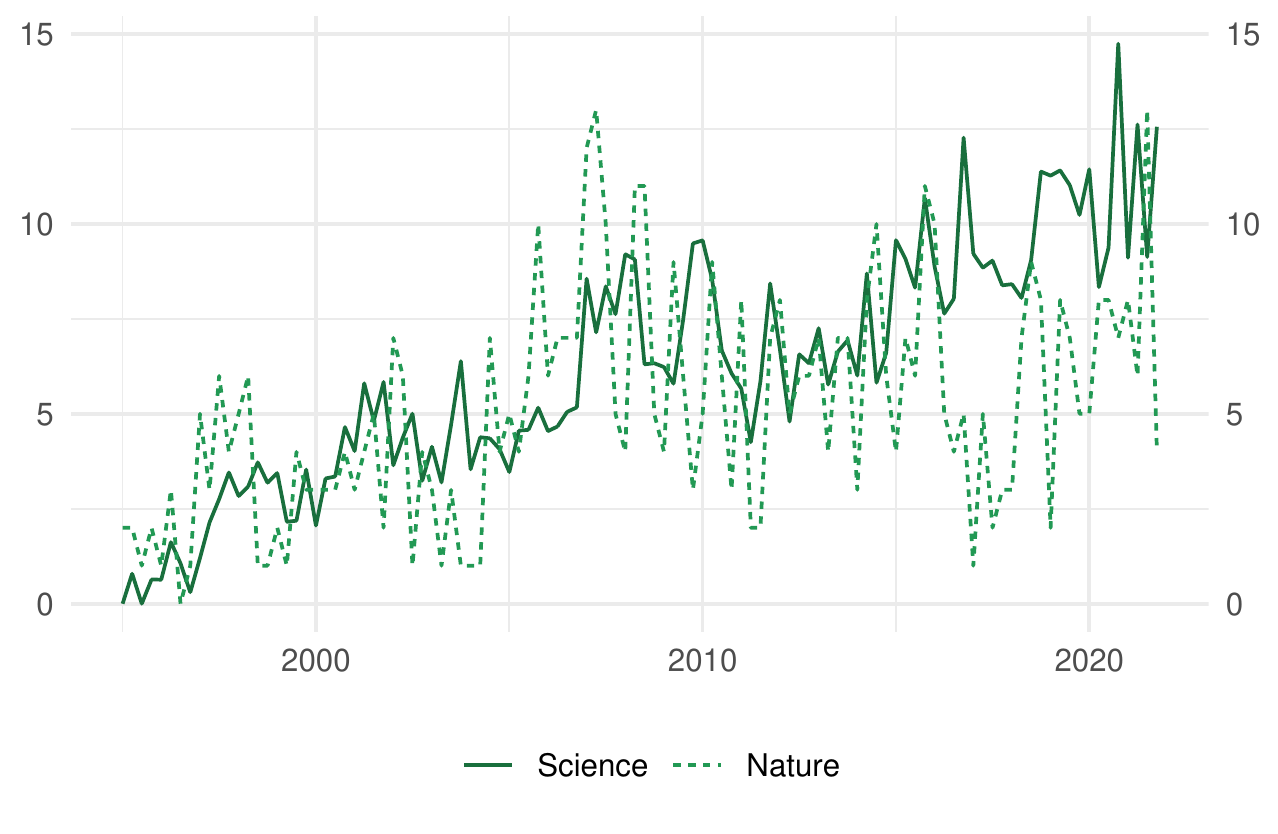}
\caption{Quarterly Share of Articles in Science containing "Climate Change".}
\label{fig:science}
\end{figure}

\section*{Appendix C}

\begin{figure}[h!]
\centering
\includegraphics[scale=0.25]{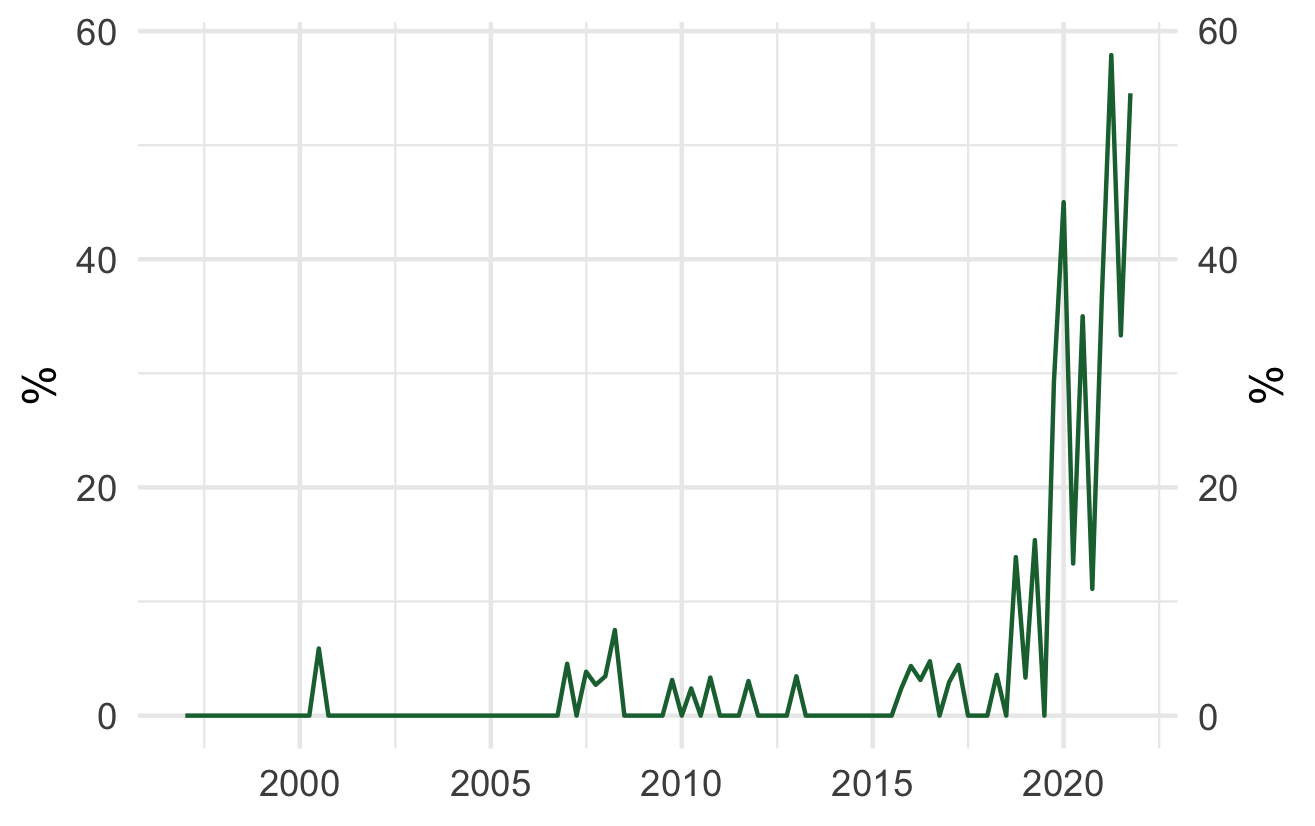}
\caption{Share of Speeches from the ECB containing the words "Climate Change"}
\label{fig:ECBspeeches}
\end{figure}

In the following table we show a comparison of mentions over time in ECB presidential speeches of climate change  with taxex and inequality. 

\newpage

\begin{table}[h!]
\centering
\begin{tabular}{rlllllll}
  \hline
 & date & n & climate  & global  & tax & taxes & inequality \\ 
  &  &  &  change &  warming &  &  &  \\   \hline
1 & 1997 Q1 & 2 & 0 & 0 & 2 & 0 & 0 \\ 
  2 & 1997 Q2 & 6 & 0 & 0 & 1 & 1 & 0 \\ 
  3 & 1997 Q3 & 2 & 0 & 0 & 0 & 0 & 0 \\ 
  4 & 1997 Q4 & 9 & 0 & 0 & 4 & 1 & 0 \\ 
  5 & 1998 Q1 & 7 & 0 & 0 & 2 & 1 & 0 \\ 
  6 & 1998 Q2 & 1 & 0 & 0 & 0 & 0 & 0 \\ 
  7 & 1998 Q3 & 8 & 0 & 0 & 1 & 1 & 0 \\ 
  8 & 1998 Q4 & 22 & 0 & 0 & 9 & 6 & 0 \\ 
  9 & 1999 Q1 & 20 & 0 & 0 & 8 & 4 & 0 \\ 
  10 & 1999 Q2 & 27 & 0 & 1 & 14 & 3 & 0 \\ 
  11 & 1999 Q3 & 18 & 0 & 0 & 7 & 3 & 0 \\ 
  12 & 1999 Q4 & 27 & 0 & 0 & 11 & 3 & 0 \\ 
  13 & 2000 Q1 & 14 & 0 & 0 & 7 & 2 & 0 \\ 
  14 & 2000 Q2 & 18 & 0 & 0 & 8 & 3 & 0 \\ 
  15 & 2000 Q3 & 17 & 1 & 0 & 8 & 2 & 0 \\ 
  16 & 2000 Q4 & 21 & 0 & 0 & 8 & 2 & 0 \\ 
  17 & 2001 Q1 & 14 & 0 & 0 & 9 & 2 & 1 \\ 
  18 & 2001 Q2 & 16 & 0 & 0 & 8 & 3 & 0 \\ 
  19 & 2001 Q3 & 13 & 0 & 0 & 3 & 0 & 0 \\ 
  20 & 2001 Q4 & 22 & 0 & 0 & 4 & 1 & 0 \\ 
  21 & 2002 Q1 & 20 & 0 & 1 & 9 & 4 & 0 \\ 
  22 & 2002 Q2 & 18 & 0 & 0 & 6 & 1 & 0 \\ 
  23 & 2002 Q3 & 8 & 0 & 0 & 3 & 2 & 0 \\ 
  24 & 2002 Q4 & 19 & 0 & 0 & 5 & 1 & 0 \\ 
  25 & 2003 Q1 & 12 & 0 & 0 & 5 & 3 & 0 \\ 
  26 & 2003 Q2 & 18 & 0 & 0 & 5 & 3 & 0 \\ 
  27 & 2003 Q3 & 10 & 0 & 0 & 2 & 0 & 0 \\ 
  28 & 2003 Q4 & 24 & 0 & 0 & 7 & 3 & 0 \\ 
  29 & 2004 Q1 & 16 & 0 & 0 & 9 & 6 & 0 \\ 
  30 & 2004 Q2 & 31 & 0 & 0 & 18 & 9 & 0 \\

   \hline
\end{tabular}
\end{table}
\newpage

\begin{table}[ht]
\centering
\begin{tabular}{rlllllll}
  \hline
 & date & n & climate  & global  & tax & taxes & inequality \\ 
  &  &  &  change &  warming &  &  &  \\   \hline
    31 & 2004 Q3 & 14 & 0 & 0 & 6 & 6 & 0 \\ 
  32 & 2004 Q4 & 30 & 0 & 0 & 12 & 7 & 0 \\ 
  33 & 2005 Q1 & 13 & 0 & 0 & 7 & 2 & 0 \\ 
 34 & 2005 Q2 & 29 & 0 & 0 & 13 & 9 & 1 \\ 
  35 & 2005 Q3 & 13 & 0 & 0 & 5 & 4 & 0 \\ 
  36 & 2005 Q4 & 26 & 0 & 0 & 8 & 5 & 0 \\ 
  37 & 2006 Q1 & 20 & 0 & 0 & 12 & 5 & 0 \\ 
  38 & 2006 Q2 & 31 & 0 & 0 & 16 & 7 & 0 \\ 
  39 & 2006 Q3 & 16 & 0 & 0 & 9 & 5 & 0 \\ 
  40 & 2006 Q4 & 29 & 0 & 0 & 14 & 7 & 0 \\ 
  41 & 2007 Q1 & 22 & 1 & 0 & 7 & 3 & 1 \\ 
  42 & 2007 Q2 & 33 & 0 & 0 & 8 & 4 & 2 \\ 
  43 & 2007 Q3 & 26 & 1 & 0 & 6 & 3 & 0 \\ 
  44 & 2007 Q4 & 37 & 1 & 0 & 14 & 4 & 1 \\ 
  45 & 2008 Q1 & 29 & 1 & 0 & 7 & 6 & 0 \\ 
  46 & 2008 Q2 & 40 & 3 & 0 & 10 & 6 & 0 \\ 
  47 & 2008 Q3 & 29 & 0 & 0 & 11 & 4 & 1 \\ 
  48 & 2008 Q4 & 34 & 0 & 0 & 13 & 3 & 2 \\ 
  49 & 2009 Q1 & 26 & 0 & 0 & 7 & 1 & 0 \\ 
  50 & 2009 Q2 & 34 & 0 & 0 & 5 & 2 & 0 \\ 
  51 & 2009 Q3 & 20 & 0 & 0 & 6 & 1 & 0 \\ 
  52 & 2009 Q4 & 32 & 1 & 1 & 7 & 0 & 0 \\ 
  53 & 2010 Q1 & 22 & 0 & 0 & 7 & 2 & 0 \\ 
  54 & 2010 Q2 & 42 & 1 & 0 & 14 & 2 & 0 \\ 
  55 & 2010 Q3 & 24 & 0 & 0 & 10 & 4 & 1 \\ 
  56 & 2010 Q4 & 30 & 1 & 0 & 6 & 1 & 4 \\ 
  57 & 2011 Q1 & 28 & 0 & 0 & 11 & 3 & 1 \\ 
  58 & 2011 Q2 & 45 & 0 & 0 & 11 & 2 & 1 \\ 
  59 & 2011 Q3 & 13 & 0 & 0 & 5 & 0 & 1 \\ 
  60 & 2011 Q4 & 33 & 1 & 0 & 13 & 3 & 0 \\ 

   \hline
\end{tabular}
\end{table}

\newpage

\begin{table}[ht]
\centering
\begin{tabular}{rlllllll}
  \hline
 & date & n & climate  & global  & tax & taxes & inequality \\ 
  &  &  &  change &  warming &  &  &  \\ 

  \hline
   61 & 2012 Q1 & 11 & 0 & 0 & 5 & 0 & 0 \\ 
  62 & 2012 Q2 & 33 & 0 & 0 & 13 & 5 & 0 \\ 
  63 & 2012 Q3 & 17 & 0 & 0 & 6 & 1 & 0 \\ 
  64 & 2012 Q4 & 30 & 0 & 0 & 17 & 4 & 1 \\ 
  65 & 2013 Q1 & 29 & 1 & 0 & 12 & 5 & 0 \\ 
  66 & 2013 Q2 & 42 & 0 & 0 & 22 & 3 & 3 \\ 
  67 & 2013 Q3 & 25 & 0 & 0 & 10 & 1 & 1 \\ 
  68 & 2013 Q4 & 37 & 0 & 0 & 20 & 4 & 3 \\ 
  69 & 2014 Q1 & 20 & 0 & 0 & 6 & 3 & 1 \\ 
  70 & 2014 Q2 & 31 & 0 & 0 & 14 & 3 & 1 \\ 
  71 & 2014 Q3 & 23 & 0 & 0 & 10 & 3 & 0 \\ 
  72 & 2014 Q4 & 30 & 0 & 0 & 9 & 4 & 1 \\ 
  73 & 2015 Q1 & 20 & 0 & 0 & 6 & 1 & 1 \\ 
  74 & 2015 Q2 & 25 & 0 & 0 & 12 & 3 & 3 \\ 
  75 & 2015 Q3 & 17 & 0 & 0 & 5 & 0 & 1 \\ 
  76 & 2015 Q4 & 42 & 1 & 0 & 12 & 3 & 1 \\ 
  77 & 2016 Q1 & 23 & 1 & 0 & 9 & 5 & 0 \\ 
  78 & 2016 Q2 & 32 & 1 & 0 & 12 & 1 & 3 \\ 
  79 & 2016 Q3 & 21 & 1 & 0 & 10 & 1 & 1 \\ 
  80 & 2016 Q4 & 34 & 0 & 0 & 19 & 2 & 4 \\ 
  81 & 2017 Q1 & 34 & 1 & 0 & 9 & 1 & 2 \\ 
  82 & 2017 Q2 & 45 & 2 & 0 & 10 & 3 & 1 \\ 
  83 & 2017 Q3 & 30 & 0 & 0 & 6 & 1 & 4 \\ 
  84 & 2017 Q4 & 39 & 0 & 0 & 10 & 2 & 4 \\ 
  85 & 2018 Q1 & 29 & 0 & 0 & 8 & 1 & 2 \\ 
  86 & 2018 Q2 & 28 & 1 & 0 & 13 & 8 & 4 \\ 
  87 & 2018 Q3 & 20 & 0 & 0 & 5 & 0 & 1 \\ 
  88 & 2018 Q4 & 36 & 5 & 1 & 11 & 1 & 2 \\ 
  89 & 2019 Q1 & 30 & 1 & 0 & 9 & 1 & 3 \\ 
  90 & 2019 Q2 & 26 & 4 & 0 & 8 & 1 & 2 \\ 
   \hline
\end{tabular}
\end{table}

\newpage

\begin{table}[ht]
\centering
\begin{tabular}{rlllllll}
  \hline
 & date & n & climate  & global  & tax & taxes & inequality \\ 
  &  &  &  change &  warming &  &  &  \\ 

  \hline
  91 & 2019 Q3 & 17 & 0 & 0 & 4 & 0 & 1 \\ 
  92 & 2019 Q4 & 34 & 10 & 1 & 13 & 6 & 6 \\ 
  93 & 2020 Q1 & 20 & 9 & 1 & 8 & 2 & 3 \\ 
  94 & 2020 Q2 & 15 & 2 & 0 & 1 & 0 & 0 \\ 
  95 & 2020 Q3 & 20 & 7 & 2 & 6 & 0 & 3 \\ 
  96 & 2020 Q4 & 27 & 3 & 0 & 6 & 0 & 3 \\ 
  97 & 2021 Q1 & 19 & 7 & 0 & 4 & 2 & 1 \\ 
  98 & 2021 Q2 & 19 & 11 & 4 & 5 & 0 & 2 \\ 
  99 & 2021 Q3 & 12 & 4 & 1 & 4 & 1 & 3 \\ 
  100 & 2021 Q4 & 11 & 8 & 2 & 3 & 0 & 1 \\ 
   \hline
\end{tabular}
\end{table}

\begin{table}[ht]
\centering
\begin{tabular}{lcccccc}
  \hline
\hline
date & n & climate change & covid & price & taxes & inequality \\ 
  \hline
1997-02-01 & 1 & 0 & 0 & 1 & 0 & 0 \\ 
  1997-03-01 & 1 & 0 & 0 & 1 & 0 & 0 \\ 
  1997-04-01 & 2 & 0 & 0 & 2 & 0 & 0 \\ 
  1997-05-01 & 1 & 0 & 0 & 1 & 0 & 0 \\ 
  1997-06-01 & 3 & 0 & 0 & 2 & 1 & 0 \\ 
  1997-09-01 & 2 & 0 & 0 & 2 & 0 & 0 \\ 
  1997-10-01 & 4 & 0 & 0 & 4 & 1 & 0 \\ 
  1997-11-01 & 5 & 0 & 0 & 5 & 0 & 0 \\ 
  1998-01-01 & 4 & 0 & 0 & 4 & 1 & 0 \\ 
  1998-02-01 & 2 & 0 & 0 & 2 & 0 & 0 \\ 
  1998-03-01 & 1 & 0 & 0 & 1 & 0 & 0 \\ 
  1998-06-01 & 1 & 0 & 0 & 1 & 0 & 0 \\ 
  1998-07-01 & 3 & 0 & 0 & 3 & 0 & 0 \\ 
  1998-09-01 & 5 & 0 & 0 & 4 & 1 & 0 \\ 
  1998-10-01 & 4 & 0 & 0 & 4 & 1 & 0 \\ 
  1998-11-01 & 9 & 0 & 0 & 9 & 2 & 0 \\ 
  1998-12-01 & 9 & 0 & 0 & 8 & 3 & 0 \\ 
  1999-01-01 & 6 & 0 & 0 & 6 & 0 & 0 \\ 
  1999-02-01 & 7 & 0 & 0 & 7 & 2 & 0 \\ 
  1999-03-01 & 7 & 0 & 0 & 7 & 2 & 0 \\ 
  1999-04-01 & 7 & 0 & 0 & 6 & 0 & 0 \\ 
  1999-05-01 & 10 & 0 & 0 & 10 & 2 & 0 \\ 
  1999-06-01 & 10 & 0 & 0 & 9 & 1 & 0 \\ 
  1999-07-01 & 5 & 0 & 0 & 5 & 1 & 0 \\ 
  1999-08-01 & 2 & 0 & 0 & 2 & 1 & 0 \\ 
  1999-09-01 & 11 & 0 & 0 & 11 & 1 & 0 \\ 
  1999-10-01 & 7 & 0 & 0 & 7 & 0 & 0 \\ 
  1999-11-01 & 16 & 0 & 0 & 14 & 3 & 0 \\ 
  1999-12-01 & 4 & 0 & 0 & 4 & 0 & 0 \\ 
     \hline
\end{tabular}
\caption{\ ECB Speeches. 1997-2022} 
\end{table}

\begin{table}[ht]
\centering
\begin{tabular}{lcccccc}
  \hline
\hline
date & n & climate change & covid & price & taxes & inequality \\ 
  \hline
  2000-01-01 & 5 & 0 & 0 & 5 & 1 & 0 \\ 
  2000-02-01 & 4 & 0 & 0 & 3 & 0 & 0 \\ 
  2000-03-01 & 5 & 0 & 0 & 5 & 1 & 0 \\ 
  2000-04-01 & 4 & 0 & 0 & 3 & 0 & 0 \\ 
  2000-05-01 & 7 & 0 & 0 & 6 & 2 & 0 \\ 
  2000-06-01 & 7 & 0 & 0 & 6 & 1 & 0 \\ 
  2000-07-01 & 1 & 0 & 0 & 1 & 0 & 0 \\ 
  2000-08-01 & 1 & 0 & 0 & 1 & 0 & 0 \\ 
  2000-09-01 & 15 & 1 & 0 & 15 & 2 & 0 \\ 
  2000-10-01 & 6 & 0 & 0 & 6 & 0 & 0 \\ 
  2000-11-01 & 10 & 0 & 0 & 10 & 1 & 0 \\ 
  2000-12-01 & 5 & 0 & 0 & 4 & 1 & 0 \\ 
  2001-01-01 & 5 & 0 & 0 & 5 & 0 & 0 \\ 
  2001-02-01 & 6 & 0 & 0 & 6 & 1 & 0 \\ 
  2001-03-01 & 3 & 0 & 0 & 2 & 1 & 1 \\ 
  2001-04-01 & 2 & 0 & 0 & 1 & 0 & 0 \\ 
  2001-05-01 & 8 & 0 & 0 & 7 & 3 & 0 \\ 
  2001-06-01 & 6 & 0 & 0 & 6 & 0 & 0 \\ 
  2001-07-01 & 1 & 0 & 0 & 1 & 0 & 0 \\ 
  2001-08-01 & 4 & 0 & 0 & 2 & 0 & 0 \\ 
  2001-09-01 & 8 & 0 & 0 & 8 & 0 & 0 \\ 
  2001-10-01 & 8 & 0 & 0 & 8 & 0 & 0 \\ 
  2001-11-01 & 9 & 0 & 0 & 8 & 0 & 0 \\ 
  2001-12-01 & 5 & 0 & 0 & 5 & 1 & 0 \\ 
  2002-01-01 & 4 & 0 & 0 & 4 & 1 & 0 \\ 
  2002-02-01 & 7 & 0 & 0 & 7 & 2 & 0 \\ 
  2002-03-01 & 9 & 0 & 0 & 8 & 1 & 0 \\ 
  2002-04-01 & 7 & 0 & 0 & 6 & 0 & 0 \\ 
  2002-05-01 & 7 & 0 & 0 & 6 & 1 & 0 \\ 
  2002-06-01 & 4 & 0 & 0 & 2 & 0 & 0 \\ 
       \hline
\end{tabular}
\caption{\ ECB Speeches. 1997-2022} 
\end{table}

\begin{table}[ht]
\centering
\begin{tabular}{lcccccc}
  \hline
\hline
date & n & climate change & covid & price & taxes & inequality \\ 
  \hline
  2002-07-01 & 5 & 0 & 0 & 4 & 1 & 0 \\ 
  2002-08-01 & 1 & 0 & 0 & 1 & 0 & 0 \\ 
  2002-09-01 & 2 & 0 & 0 & 1 & 1 & 0 \\ 
  2002-10-01 & 5 & 0 & 0 & 5 & 0 & 0 \\ 
  2002-11-01 & 8 & 0 & 0 & 7 & 0 & 0 \\ 
  2002-12-01 & 6 & 0 & 0 & 5 & 1 & 0 \\ 
  2003-01-01 & 1 & 0 & 0 & 1 & 0 & 0 \\ 
  2003-02-01 & 5 & 0 & 0 & 5 & 3 & 0 \\ 
  2003-03-01 & 6 & 0 & 0 & 5 & 0 & 0 \\ 
  2003-04-01 & 5 & 0 & 0 & 4 & 1 & 0 \\ 
  2003-05-01 & 4 & 0 & 0 & 4 & 0 & 0 \\ 
  2003-06-01 & 9 & 0 & 0 & 8 & 2 & 0 \\ 
  2003-07-01 & 5 & 0 & 0 & 5 & 0 & 0 \\ 
  2003-08-01 & 1 & 0 & 0 & 1 & 0 & 0 \\ 
  2003-09-01 & 4 & 0 & 0 & 4 & 0 & 0 \\ 
  2003-10-01 & 9 & 0 & 0 & 6 & 0 & 0 \\ 
  2003-11-01 & 12 & 0 & 0 & 10 & 2 & 0 \\ 
  2003-12-01 & 3 & 0 & 0 & 2 & 1 & 0 \\ 
  2004-01-01 & 6 & 0 & 0 & 4 & 3 & 0 \\ 
  2004-02-01 & 6 & 0 & 0 & 5 & 2 & 0 \\ 
  2004-03-01 & 4 & 0 & 0 & 4 & 1 & 0 \\ 
  2004-04-01 & 10 & 0 & 0 & 9 & 5 & 0 \\ 
  2004-05-01 & 13 & 0 & 0 & 12 & 3 & 0 \\ 
  2004-06-01 & 8 & 0 & 0 & 5 & 1 & 0 \\ 
  2004-07-01 & 3 & 0 & 0 & 3 & 3 & 0 \\ 
  2004-08-01 & 1 & 0 & 0 & 1 & 0 & 0 \\ 
  2004-09-01 & 10 & 0 & 0 & 8 & 3 & 0 \\ 
  2004-10-01 & 11 & 0 & 0 & 9 & 2 & 0 \\ 
  2004-11-01 & 10 & 0 & 0 & 8 & 3 & 0 \\ 
  2004-12-01 & 9 & 0 & 0 & 8 & 2 & 0 \\
         \hline
\end{tabular}
\caption{\ ECB Speeches. 1997-2022} 
\end{table}

\begin{table}[ht]
\centering
\begin{tabular}{lcccccc}
  \hline
\hline
date & n & climate change & covid & price & taxes & inequality \\ 
  \hline
  2005-01-01 & 5 & 0 & 0 & 3 & 1 & 0 \\ 
  2005-02-01 & 2 & 0 & 0 & 1 & 0 & 0 \\ 
  2005-03-01 & 6 & 0 & 0 & 6 & 1 & 0 \\ 
  2005-04-01 & 7 & 0 & 0 & 6 & 2 & 0 \\ 
  2005-05-01 & 10 & 0 & 0 & 9 & 5 & 0 \\ 
  2005-06-01 & 12 & 0 & 0 & 12 & 2 & 1 \\ 
  2005-07-01 & 5 & 0 & 0 & 4 & 1 & 0 \\ 
  2005-08-01 & 2 & 0 & 0 & 1 & 0 & 0 \\ 
  2005-09-01 & 6 & 0 & 0 & 5 & 3 & 0 \\ 
  2005-10-01 & 12 & 0 & 0 & 9 & 2 & 0 \\ 
  2005-11-01 & 11 & 0 & 0 & 9 & 3 & 0 \\ 
  2005-12-01 & 3 & 0 & 0 & 3 & 0 & 0 \\ 
  2006-01-01 & 3 & 0 & 0 & 2 & 0 & 0 \\ 
  2006-02-01 & 8 & 0 & 0 & 6 & 4 & 0 \\ 
  2006-03-01 & 9 & 0 & 0 & 7 & 1 & 0 \\ 
  2006-04-01 & 6 & 0 & 0 & 6 & 1 & 0 \\ 
  2006-05-01 & 13 & 0 & 0 & 12 & 4 & 0 \\ 
  2006-06-01 & 12 & 0 & 0 & 11 & 2 & 0 \\ 
  2006-07-01 & 6 & 0 & 0 & 4 & 1 & 0 \\ 
  2006-09-01 & 10 & 0 & 0 & 8 & 4 & 0 \\ 
  2006-10-01 & 11 & 0 & 0 & 10 & 4 & 0 \\ 
  2006-11-01 & 12 & 0 & 0 & 11 & 1 & 0 \\ 
  2006-12-01 & 6 & 0 & 0 & 6 & 2 & 0 \\ 
  2007-01-01 & 8 & 0 & 0 & 7 & 1 & 1 \\ 
  2007-02-01 & 6 & 0 & 0 & 4 & 0 & 0 \\ 
  2007-03-01 & 8 & 1 & 0 & 6 & 2 & 0 \\ 
  2007-04-01 & 6 & 0 & 0 & 6 & 1 & 0 \\ 
  2007-05-01 & 12 & 0 & 0 & 9 & 0 & 2 \\ 
  2007-06-01 & 15 & 0 & 0 & 14 & 3 & 0 \\ 
           \hline
\end{tabular}
\caption{\ ECB Speeches. 1997-2022} 
\end{table}
  
\begin{table}[ht]
\centering
\begin{tabular}{lcccccc}
  \hline
\hline
date & n & climate change & covid & price & taxes & inequality \\ 
  \hline
  2007-07-01 & 8 & 0 & 0 & 8 & 1 & 0 \\ 
  2007-08-01 & 1 & 0 & 0 & 1 & 1 & 0 \\ 
  2007-09-01 & 17 & 1 & 0 & 15 & 1 & 0 \\ 
  2007-10-01 & 14 & 0 & 0 & 14 & 2 & 0 \\ 
  2007-11-01 & 14 & 1 & 0 & 13 & 1 & 0 \\ 
  2007-12-01 & 9 & 0 & 0 & 8 & 1 & 1 \\ 
  2008-01-01 & 13 & 0 & 0 & 10 & 3 & 0 \\ 
  2008-02-01 & 9 & 1 & 0 & 7 & 1 & 0 \\ 
  2008-03-01 & 7 & 0 & 0 & 6 & 2 & 0 \\ 
  2008-04-01 & 18 & 0 & 0 & 17 & 3 & 0 \\ 
  2008-05-01 & 8 & 1 & 0 & 6 & 2 & 0 \\ 
  2008-06-01 & 14 & 2 & 0 & 14 & 1 & 0 \\ 
  2008-07-01 & 5 & 0 & 0 & 4 & 2 & 0 \\ 
  2008-08-01 & 1 & 0 & 0 & 1 & 0 & 0 \\ 
  2008-09-01 & 23 & 0 & 0 & 18 & 2 & 1 \\ 
  2008-10-01 & 9 & 0 & 0 & 8 & 1 & 1 \\ 
  2008-11-01 & 16 & 0 & 0 & 14 & 2 & 1 \\ 
  2008-12-01 & 9 & 0 & 0 & 9 & 0 & 0 \\ 
  2009-01-01 & 7 & 0 & 0 & 6 & 0 & 0 \\ 
  2009-02-01 & 10 & 0 & 0 & 7 & 1 & 0 \\ 
  2009-03-01 & 9 & 0 & 0 & 9 & 0 & 0 \\ 
  2009-04-01 & 7 & 0 & 0 & 6 & 0 & 0 \\ 
  2009-05-01 & 8 & 0 & 0 & 6 & 1 & 0 \\ 
  2009-06-01 & 19 & 0 & 0 & 17 & 1 & 0 \\ 
  2009-07-01 & 4 & 0 & 0 & 1 & 0 & 0 \\ 
  2009-08-01 & 1 & 0 & 0 & 1 & 0 & 0 \\ 
  2009-09-01 & 15 & 0 & 0 & 12 & 1 & 0 \\ 
  2009-10-01 & 9 & 0 & 0 & 8 & 0 & 0 \\ 
  2009-11-01 & 14 & 0 & 0 & 11 & 0 & 0 \\ 
  2009-12-01 & 9 & 1 & 0 & 6 & 0 & 0 \\ 
             \hline
\end{tabular}
\caption{\ ECB Speeches. 1997-2022} 
\end{table}

\begin{table}[ht]
\centering
\begin{tabular}{lcccccc}
  \hline
\hline
date & n & climate change & covid & price & taxes & inequality \\ 
  \hline
  2010-01-01 & 5 & 0 & 0 & 4 & 0 & 0 \\ 
  2010-02-01 & 7 & 0 & 0 & 6 & 1 & 0 \\ 
  2010-03-01 & 10 & 0 & 0 & 7 & 1 & 0 \\ 
  2010-04-01 & 16 & 1 & 0 & 13 & 1 & 0 \\ 
  2010-05-01 & 12 & 0 & 0 & 9 & 0 & 0 \\ 
  2010-06-01 & 14 & 0 & 0 & 12 & 1 & 0 \\ 
  2010-07-01 & 6 & 0 & 0 & 5 & 2 & 0 \\ 
  2010-08-01 & 1 & 0 & 0 & 1 & 1 & 1 \\ 
  2010-09-01 & 17 & 0 & 0 & 10 & 1 & 0 \\ 
  2010-10-01 & 12 & 0 & 0 & 10 & 0 & 2 \\ 
  2010-11-01 & 14 & 0 & 0 & 12 & 1 & 1 \\ 
  2010-12-01 & 4 & 1 & 0 & 4 & 0 & 1 \\ 
  2011-01-01 & 6 & 0 & 0 & 6 & 0 & 0 \\ 
  2011-02-01 & 10 & 0 & 0 & 8 & 0 & 0 \\ 
  2011-03-01 & 12 & 0 & 0 & 7 & 3 & 1 \\ 
  2011-04-01 & 5 & 0 & 0 & 4 & 0 & 0 \\ 
  2011-05-01 & 20 & 0 & 0 & 17 & 1 & 1 \\ 
  2011-06-01 & 20 & 0 & 0 & 17 & 1 & 0 \\ 
  2011-07-01 & 3 & 0 & 0 & 0 & 0 & 0 \\ 
  2011-08-01 & 3 & 0 & 0 & 2 & 0 & 1 \\ 
  2011-09-01 & 7 & 0 & 0 & 5 & 0 & 0 \\ 
  2011-10-01 & 14 & 1 & 0 & 13 & 1 & 0 \\ 
  2011-11-01 & 12 & 0 & 0 & 10 & 0 & 0 \\ 
  2011-12-01 & 7 & 0 & 0 & 7 & 2 & 0 \\ 
  2012-02-01 & 4 & 0 & 0 & 4 & 0 & 0 \\ 
  2012-03-01 & 7 & 0 & 0 & 5 & 0 & 0 \\ 
  2012-04-01 & 12 & 0 & 0 & 11 & 4 & 0 \\ 
  2012-05-01 & 11 & 0 & 0 & 9 & 1 & 0 \\ 
  2012-06-01 & 10 & 0 & 0 & 8 & 0 & 0 \\ 
               \hline
\end{tabular}
\caption{\ ECB Speeches. 1997-2022} 
\end{table}

\begin{table}[ht]
\centering
\begin{tabular}{lcccccc}
  \hline
\hline
date & n & climate change & covid & price & taxes & inequality \\ 
  \hline
  2012-07-01 & 7 & 0 & 0 & 2 & 1 & 0 \\ 
  2012-08-01 & 2 & 0 & 0 & 2 & 0 & 0 \\ 
  2012-09-01 & 8 & 0 & 0 & 7 & 0 & 0 \\ 
  2012-10-01 & 12 & 0 & 0 & 10 & 3 & 1 \\ 
  2012-11-01 & 11 & 0 & 0 & 9 & 1 & 0 \\ 
  2012-12-01 & 7 & 0 & 0 & 5 & 0 & 0 \\ 
  2013-01-01 & 9 & 0 & 0 & 7 & 1 & 0 \\ 
  2013-02-01 & 11 & 1 & 0 & 10 & 3 & 0 \\ 
  2013-03-01 & 9 & 0 & 0 & 4 & 1 & 0 \\ 
  2013-04-01 & 14 & 0 & 0 & 10 & 2 & 1 \\ 
  2013-05-01 & 13 & 0 & 0 & 9 & 1 & 0 \\ 
  2013-06-01 & 15 & 0 & 0 & 13 & 0 & 2 \\ 
  2013-07-01 & 9 & 0 & 0 & 5 & 1 & 0 \\ 
  2013-08-01 & 2 & 0 & 0 & 2 & 0 & 0 \\ 
  2013-09-01 & 14 & 0 & 0 & 12 & 0 & 1 \\ 
  2013-10-01 & 12 & 0 & 0 & 10 & 3 & 3 \\ 
  2013-11-01 & 17 & 0 & 0 & 11 & 0 & 0 \\ 
  2013-12-01 & 8 & 0 & 0 & 7 & 1 & 0 \\ 
  2014-01-01 & 7 & 0 & 0 & 4 & 1 & 0 \\ 
  2014-02-01 & 7 & 0 & 0 & 6 & 1 & 1 \\ 
  2014-03-01 & 6 & 0 & 0 & 5 & 1 & 0 \\ 
  2014-04-01 & 11 & 0 & 0 & 11 & 0 & 1 \\ 
  2014-05-01 & 14 & 0 & 0 & 12 & 1 & 0 \\ 
  2014-06-01 & 6 & 0 & 0 & 6 & 2 & 0 \\ 
  2014-07-01 & 9 & 0 & 0 & 8 & 1 & 0 \\ 
  2014-08-01 & 1 & 0 & 0 & 1 & 1 & 0 \\ 
  2014-09-01 & 13 & 0 & 0 & 9 & 1 & 0 \\ 
  2014-10-01 & 12 & 0 & 0 & 9 & 1 & 1 \\ 
  2014-11-01 & 15 & 0 & 0 & 11 & 2 & 0 \\ 
  2014-12-01 & 3 & 0 & 0 & 3 & 1 & 0 \\ 
                 \hline
\end{tabular}
\caption{\ ECB Speeches. 1997-2022} 
\end{table}

\begin{table}[ht]
\centering
\begin{tabular}{lcccccc}
  \hline
\hline
date & n & climate change & covid & price & taxes & inequality \\ 
  \hline
  2015-01-01 & 2 & 0 & 0 & 1 & 0 & 0 \\ 
  2015-02-01 & 6 & 0 & 0 & 4 & 0 & 0 \\ 
  2015-03-01 & 12 & 0 & 0 & 10 & 1 & 1 \\ 
  2015-04-01 & 9 & 0 & 0 & 8 & 1 & 0 \\ 
  2015-05-01 & 8 & 0 & 0 & 8 & 1 & 1 \\ 
  2015-06-01 & 8 & 0 & 0 & 5 & 1 & 2 \\ 
  2015-07-01 & 4 & 0 & 0 & 2 & 0 & 0 \\ 
  2015-08-01 & 3 & 0 & 0 & 3 & 0 & 1 \\ 
  2015-09-01 & 10 & 0 & 0 & 4 & 0 & 0 \\ 
  2015-10-01 & 16 & 1 & 0 & 10 & 0 & 0 \\ 
  2015-11-01 & 22 & 0 & 0 & 11 & 3 & 1 \\ 
  2015-12-01 & 4 & 0 & 0 & 3 & 0 & 0 \\ 
  2016-01-01 & 9 & 0 & 0 & 5 & 3 & 0 \\ 
  2016-02-01 & 8 & 1 & 0 & 8 & 0 & 0 \\ 
  2016-03-01 & 6 & 0 & 0 & 5 & 2 & 0 \\ 
  2016-04-01 & 11 & 0 & 0 & 9 & 1 & 1 \\ 
  2016-05-01 & 5 & 0 & 0 & 3 & 0 & 1 \\ 
  2016-06-01 & 16 & 1 & 0 & 10 & 0 & 1 \\ 
  2016-07-01 & 6 & 0 & 0 & 5 & 0 & 0 \\ 
  2016-08-01 & 2 & 0 & 0 & 2 & 0 & 0 \\ 
  2016-09-01 & 13 & 1 & 0 & 9 & 1 & 1 \\ 
  2016-10-01 & 14 & 0 & 0 & 13 & 0 & 2 \\ 
  2016-11-01 & 18 & 0 & 0 & 15 & 2 & 2 \\ 
  2016-12-01 & 2 & 0 & 0 & 1 & 0 & 0 \\ 
  2017-01-01 & 10 & 0 & 0 & 5 & 0 & 1 \\ 
  2017-02-01 & 9 & 1 & 0 & 9 & 1 & 1 \\ 
  2017-03-01 & 15 & 0 & 0 & 9 & 0 & 0 \\ 
  2017-04-01 & 15 & 0 & 0 & 9 & 1 & 1 \\ 
  2017-05-01 & 20 & 1 & 0 & 15 & 2 & 0 \\ 
  2017-06-01 & 10 & 1 & 0 & 6 & 0 & 0 \\ 
                   \hline
\end{tabular}
\caption{\ ECB Speeches. 1997-2022} 
\end{table}

\begin{table}[ht]
\centering
\begin{tabular}{lcccccc}
  \hline
\hline
date & n & climate change & covid & price & taxes & inequality \\ 
  \hline
  2017-07-01 & 8 & 0 & 0 & 7 & 0 & 0 \\ 
  2017-08-01 & 3 & 0 & 0 & 2 & 1 & 2 \\ 
  2017-09-01 & 19 & 0 & 0 & 12 & 0 & 2 \\ 
  2017-10-01 & 15 & 0 & 0 & 10 & 0 & 2 \\ 
  2017-11-01 & 22 & 0 & 0 & 15 & 1 & 2 \\ 
  2017-12-01 & 2 & 0 & 0 & 2 & 1 & 0 \\ 
  2018-01-01 & 4 & 0 & 0 & 3 & 1 & 0 \\ 
  2018-02-01 & 15 & 0 & 0 & 10 & 0 & 2 \\ 
  2018-03-01 & 10 & 0 & 0 & 6 & 0 & 0 \\ 
  2018-04-01 & 9 & 0 & 0 & 8 & 2 & 1 \\ 
  2018-05-01 & 14 & 1 & 0 & 13 & 6 & 3 \\ 
  2018-06-01 & 5 & 0 & 0 & 5 & 0 & 0 \\ 
  2018-07-01 & 7 & 0 & 0 & 4 & 0 & 1 \\ 
  2018-08-01 & 1 & 0 & 0 & 1 & 0 & 0 \\ 
  2018-09-01 & 12 & 0 & 0 & 10 & 0 & 0 \\ 
  2018-10-01 & 11 & 2 & 0 & 11 & 0 & 1 \\ 
  2018-11-01 & 20 & 3 & 0 & 13 & 1 & 1 \\ 
  2018-12-01 & 5 & 0 & 0 & 2 & 0 & 0 \\ 
  2019-01-01 & 8 & 1 & 0 & 4 & 0 & 0 \\ 
  2019-02-01 & 11 & 0 & 0 & 8 & 0 & 1 \\ 
  2019-03-01 & 11 & 0 & 0 & 9 & 1 & 2 \\ 
  2019-04-01 & 4 & 1 & 0 & 2 & 0 & 0 \\ 
  2019-05-01 & 13 & 2 & 0 & 7 & 1 & 2 \\ 
  2019-06-01 & 9 & 1 & 0 & 5 & 0 & 0 \\ 
  2019-07-01 & 5 & 0 & 0 & 5 & 0 & 0 \\ 
  2019-08-01 & 1 & 0 & 0 & 1 & 0 & 0 \\ 
  2019-09-01 & 11 & 0 & 0 & 6 & 0 & 1 \\ 
  2019-10-01 & 10 & 3 & 0 & 9 & 3 & 3 \\ 
  2019-11-01 & 18 & 4 & 0 & 12 & 2 & 2 \\ 
  2019-12-01 & 6 & 3 & 0 & 5 & 1 & 1 \\ 
                     \hline
\end{tabular}
\caption{\ ECB Speeches. 1997-2022} 
\end{table}

\begin{table}[ht]
\centering
\begin{tabular}{lcccccc}
  \hline
\hline
date & n & climate change & covid & price & taxes & inequality \\ 
  \hline
  2020-01-01 & 4 & 2 & 0 & 1 & 0 & 1 \\ 
  2020-02-01 & 15 & 7 & 0 & 14 & 2 & 2 \\ 
  2020-03-01 & 1 & 0 & 0 & 1 & 0 & 0 \\ 
  2020-04-01 & 2 & 0 & 2 & 2 & 0 & 0 \\ 
  2020-05-01 & 5 & 1 & 3 & 4 & 0 & 0 \\ 
  2020-06-01 & 8 & 1 & 7 & 8 & 0 & 0 \\ 
  2020-07-01 & 6 & 1 & 5 & 3 & 0 & 0 \\ 
  2020-08-01 & 2 & 1 & 2 & 2 & 0 & 0 \\ 
  2020-09-01 & 12 & 5 & 10 & 8 & 0 & 3 \\ 
  2020-10-01 & 9 & 1 & 5 & 4 & 0 & 1 \\ 
  2020-11-01 & 14 & 2 & 11 & 13 & 0 & 1 \\ 
  2020-12-01 & 4 & 0 & 4 & 1 & 0 & 1 \\ 
  2021-01-01 & 5 & 3 & 3 & 3 & 1 & 1 \\ 
  2021-02-01 & 5 & 2 & 4 & 4 & 1 & 0 \\ 
  2021-03-01 & 9 & 2 & 8 & 5 & 0 & 0 \\ 
  2021-04-01 & 6 & 3 & 5 & 5 & 0 & 1 \\ 
  2021-05-01 & 4 & 3 & 1 & 3 & 0 & 1 \\ 
  2021-06-01 & 9 & 5 & 6 & 5 & 0 & 0 \\ 
  2021-07-01 & 3 & 2 & 2 & 3 & 0 & 2 \\ 
  2021-08-01 & 1 & 0 & 1 & 1 & 0 & 0 \\ 
  2021-09-01 & 8 & 2 & 4 & 7 & 1 & 1 \\ 
  2021-10-01 & 11 & 8 & 6 & 8 & 0 & 1 \\ 
  2021-11-01 & 18 & 8 & 9 & 13 & 2 & 3 \\ 
  2021-12-01 & 4 & 2 & 4 & 4 & 0 & 1 \\ 
  2022-01-01 & 3 & 2 & 2 & 3 & 1 & 0 \\ 
  2022-02-01 & 11 & 3 & 3 & 8 & 0 & 1 \\ 
  2022-03-01 & 10 & 4 & 4 & 9 & 2 & 1 \\ 
  2022-04-01 & 10 & 6 & 4 & 10 & 2 & 1 \\ 
  2022-05-01 & 6 & 0 & 2 & 6 & 0 & 1 \\ 
  2022-06-01 & 9 & 2 & 2 & 6 & 0 & 0 \\ 
   \hline
\end{tabular}
\caption{\ ECB Speeches. 1997-2022} 
\end{table}

\end{document}